\documentclass[useAMS]{mn2e}

\usepackage[ansinew]{inputenc}
\usepackage{graphicx} 

\title[Nuclear and Extended NIR Spectra of NGC1068]{Nuclear and Extended 
Spectra of NGC 1068 - I: Hints from Near-Infrared Spectroscopy}
\author[Martins et al.]{Lucimara P. Martins$^{1}$\thanks{Email:lucimara.martins@cruzeirodosul.edu.br}\footnotemark[2] , Alberto Rodr\'{\i}guez-Ardila$^{2}$\thanks{Visiting Astronomer at the Infrared Telescope Facility, which is operated
by the University of Hawaii under Cooperative Agreement no. NNX-08AE38A
with the National Aeronautics and Space Administration, Science Mission
Directorate, Planetary Astronomy Program.
}, Ronaldo de Souza$^{3}$,\newauthor and Ruth Gruenwald$^{3}$ \\
$^{1}$NAT - Universidade Cruzeiro do Sul, Rua Galvao Bueno, S\~ao Paulo, SP, Brazil. \\
$^{2}$Laborat\'orio Nacional de Astrof\'isica/MCT, Rua dos Estados Unidos 154, Bairro das Na\c c\~oes,
CEP 37504-364, Itajub\'a, MG, Brazil.\\
$^{3}$Instituto Astron\^omico e Geof\'isico - USP, Rua do Mat\~ao, 1226, S\~ao Paulo, SP}

\begin{document}

\date{Accepted ? December ? Received ? December ?; in original form ? October ?}

\pagerange{\pageref{firstpage}--\pageref{lastpage}} \pubyear{2009}

\maketitle

\label{firstpage}

\begin{abstract}

We report the first simultaneous zJHK spectroscopy on the archetypical
Seyfert~2 Galaxy NGC~1068 covering the wavelength region 0.9 to 2.4$\mu$m.
The slit, aligned in the NS direction and centred in the optical nucleus,
maps a region 300~pc in radius at sub-arcsec resolution, with a spectral
resolving power of 360~km~s$^{-1}$. This configuration allow us to study
the physical properties of the nuclear gas including that of the north
side of the ionization cone, map the strong excess of continuum emission
in the K-band and attributed to dust and study the variations, both in
flux and profile, in the emission lines. Our results show that (1) Mid- to
low-ionization emission lines are splitted into two components, whose
relative strengths vary with the position along the slit and seem to be
correlated with the jet. (2) The coronal lines are single-peaked and are
detected only in the central few hundred of parsecs from the nucleus. (3)
The absorption lines indicate the presence of intermediate age stellar
population, which might be a signicant contributor to the continuum in the
NIR spectra. (4) Through some simple photoionization models we find
photoionization as the main mechanism powering the emitting gas. (5)
Calculations using stellar features point to a mass concentration inside
the 100 - 200 pc of about 10$^{10}~M_{\odot}$.

\end{abstract}


\section{Introduction}

Near-IR spectroscopy has been playing a unique
role in our understanding of the AGN phenomenon for
several reasons. First, it includes a wealth of
emission lines and stellar absorption features not
observed in the optical, which are useful
for studying the physical properties of the emission gas and to
determine the age and metallicity of the nuclear stellar
population as, for instance, the strong CO bandheads in the H and K
bands (e.g., Schreiber 1998; Origlia, Moorwood, \& Oliva
1993). Second, extinction by dust is attenuated by a factor
of ten relative to that of the optical, allowing
to probe depths unreachable at shorter wavelengths.
Third, it is a transition region, where the continuum
from the central source no longer dominates while
the thermal continuum produced by dust and stars
becomes important.

In the last years, the interest
for observation in this interval has increased, and
thanks to the availability of cross-dispersed (XD)
spectrographs, it is possible to study the whole
0.8-2.4$\mu$m region in a single observation, avoiding the
aperture and seeing effects that usually affects $JHK$
spectroscopy done in long-slit single band observations.

With the above in  mind, here we present the first
spatially resolved XD spectroscopy covering the interval 0.8-2.4$\mu$m
made for NGC1068. This object is one of the nearest and probably the most
intensely studied Seyfert 2 galaxy. Observations in all wavelength bands
from radio to hard X-rays have formed a
uniquely detailed picture of this source. NGC 1068 has
played a unique role in the development of unified scenarios for
Seyfert 1 and Seyfert 2 galaxies (Antonucci \& Miller 1985),
in the study of molecular gas in the nuclear region of Seyferts
(e.g., Myers \& Scoville 1987; Tacconi et al. 1994), and
in elucidating the importance of star formation activity
coexistent with the active galactic nucleus (AGN), on both
larger (e.g., Telesco \& Decher 1988) and smaller (Macchetto
et al. 1994; Thatte et al. 1997) scales. NGC 1068 also hosts a
prominent narrow-line region (NLR) that is approximately
co-spatial with a linear radio source with two lobes (Wilson
\& Ulvestad 1983). The narrow emission line region has been
extensively characterized from subarcsecond clouds probed
 by the Hubble Space Telescope  (HST) (Evans et al. 1991;
Macchetto et al. 1994) to the ionization cone and extended
emission-line region (Pogge 1988; Unger et al. 1992) extending
to radii of at least 30$\arcsec$  (1$\arcsec$=72 pc at the distance  of
14.4 Mpc; Tully 1988).

From approximately 2" southwest to the nucleus to 4" northeast,
emission lines exhibit multiple components (Cecil et al. 1990;
Crenshaw \& Kraemer 2000a). Broad lines were found to be approximately
2500 - 4000 km s$^{-1}$ wide, which may be linked to those that are  found
in polarized light and believed to be reflected light from the  inner
broad line region (BLR).
Narrow lines consist of a pair of red and blue components. Optical
studies of [O\,{\sc ii}] and [N\,{\sc ii}] line profiles suggest that the  separation of
these two components varies across the conical NLR.  Between $\sim$2".5 and
4".5 northeast from the nucleus,
UV line emission is redshifted relative to the systemic value, a
pattern that is interpreted as the expansion of the plasma in the radio
lobe (Axon et al. 1998).

The mechanisms powering the gas of the NLR, i.e. photoionization
from the nucleus or shocks produced by jets, have long been under
debate. Dopita \& Sutherland (1996) and Bicknell et al. (1998) proposed 
that the emission in the NLR may be entirely caused by shocks. Velocity 
splitting over 1000 km$^{-1}$, reported by Axon et al. (1998), in the 
vicinity of some of the bright emission-line knots provides evidence  that
fast shocks exist in the NLR of NGC 1068. However, more recent HST  data
(Crenshaw \& Kraemer 2000b; Cecil et al. 2002; Mazzalay et al. 2009) 
found that the emission line ratios are consistent with photoionization 
instead of shock heating mechanism. On the other hand, models accounting 
for  both photoionization from the central radiation and shocks were 
required to explain both the continuum and emission-lines observed in the spectra of 
active galaxies (e.g., Contini, Rodr\'{\i}guez-Ardila \& Viegas 2003,
Rodr\'{\i}guez-Ardila, Contini  \& Viegas 2005).

The scenario in NGC~1068 is thus clearly complex, and observational
constraints from all wavelengths need to be put together if we want to
understand the processes taking place in this galaxy. Spatially resolved
NLR spectroscopy in the NIR can contribute to the solution of the NGC~1068
puzzle. Here, we present the first observations on this source covering
simultaneously the 0.8$-2.4~\mu$m region at moderate spectral and spatial
resolution. Our data are thus able to map a wavelength interval rich in
emission lines with a large range of ionization and absorption features 
from the stellar population not only from the nucleus but also from the
circumnuclear region, poorly studied in the literature.
In \S 2 a description of the observations is given. The main characteristics
of the nuclear and  extended spectra are presented in \S 3
and \S 4 , respectively.  Photoionization model predictions for the most 
intense lines are discussed in \S 5, whereas  the calculations of the central mass 
based on stellar features appear in \S 6. Finally, our concluding remarks
are presented in \S 7.

\section{The Observations}

The spectra were obtained at the NASA 3m Infrared
Telescope Facility (IRTF) in October 30, 2007. The
SpeX spectrograph (Rayner et al., 2003) was used in the short
cross-dispersed mode (SXD, 0.8 - 2.4 $\mu$m).  The detector employed
consisted
of a 1024x1024 ALADDIN 3 InSb array with a spatial scale
of 0.15"/pixel. A 0.8"x 15" slit oriented in the north-south direction
was used, providing a spectral
resolution of 360 km/s. This value was determined both from
the arc lamp and the sky line spectra and was found to
be constant with wavelength along the observed spectra. During the
observations the seeing was $\sim$1.3".
Observations were
done nodding in an object-sky-sky-object  pattern with integration
time of 180 s per frame and total on-source integration
time of 12 minutes.  After the galaxy, the A0V star HD18571 was
observed as telluric standard  and to flux calibrate the object.
The spectral reduction, extraction and wavelength calibration
procedures were performed using SPEXTOOL, the in-house
software developed and provided by the SpeX team for
the IRTF community (Cushing, Vacca \& Rayner, 2004). 
The S/N of the spectra varies from $\sim$43 at the nuclear aperture to
$\sim$13 at the most external apertures. Intermediate regions have S/N of
$\sim$20.

Figure~1 shows the position of the slit superimposed
in the galaxy contours obtained from Galliano et al. 
(2003).  From our data, we found that the light peak of the nuclear 
aperture in the K-band coincides in position with that determined
by Galliano et al. (2003) for the same band.
The gray contours show the 6 cm emission
(Gallimore et al. 1996) and the red dotted contours
show the 20 $\micron$ image (Alloin et al. 2000).
For NGC~1068, 17 extractions were made along the spatial
direction:  one, of radius 0.8'', 
centered at the peak of the K-band
distribution, and eight
more, of 0.2'' in radius, at each side of it. In addition, in order
to study in detail the variations of line profiles and continuum 
emission with distance, the nuclear aperture were subdivided into seven
smaller apertures (called nuc$\_$a to nuc$\_$g in Figure~1) of 0.2'' of
radius each. We are aware that these smaller apertures cannot
be used for flux estimates because they are oversampled.
The 1-D spectra were then corrected for telluric absorption
and flux calibrated using Xtellcor (Vacca, Cushing \& Rayner,
2003), another in-house software developed by the IRTF team.
Finally, the different orders of the galaxy spectrum were
merged to form a single 1-D frame. It was later corrected
for redshift, determined from the average z  measured from
the most conspicuous lines. No Galactic extinction correction was
applied because it is negligible at NIR (A$_{\rm nir}$=0.112).

\begin{figure} 
\includegraphics [width=88mm]{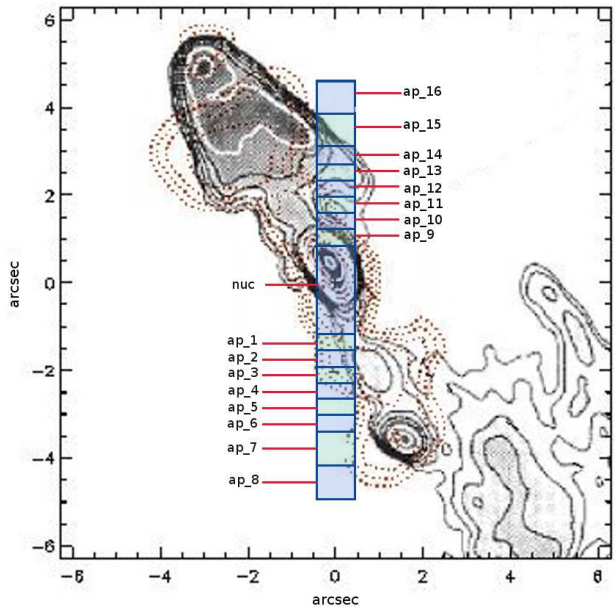}
\caption{NGC 1068 contours obtained from Galliano et al. (2003) with the 
position of the IRTF slit superimposed. The gray contours show the 6 cm emission 
(Gallimore et al. 1996) and the red dotted contours show the 20 $\micron$ image
(Alloin et al. 2000)}.
\end{figure}

\section{Nuclear Spectra}

Below we describe in detail the main features, both in emission and 
absorption, detected in the nuclear region of NGC~1068, which correspond to the 
spectrum covering 1$"$.6, centered on the maximum of the galaxy profile. The 
resulting nuclear spectra in the ZJ, H and K ranges are plotted in Figures 2, 3 and 4, 
respectively, where the wavelengths have been
translated to the observer's rest frame.

\subsection{Emission lines}

An inspection to Figures 2, 3 and 4 allows us to state that the most conspicuous 
features in the nuclear spectrum are [S\,{\sc iii}]$\lambda$$\lambda$0.907,0.953~$\micron$, 
He\,{\sc i}$\lambda$1.083~$\micron$, 
Pa$\beta$, [Fe\,{\sc ii}]$\lambda$1.643~$\micron$, Pa$\alpha$, and the coronal lines
[Si\,{\sc x}]$\lambda$1.430~$\micron$, and [Si\,{\sc vi}]$\lambda$1.962~$\micron$. 
Pa$\gamma$ and Br$\gamma$ are also present, but very weak. 
Pa$\gamma$ is in the wing of the
very strong He I line. Pa$\alpha$ could
not be measured because it partially falls in the gap between orders. 
Emission from H$_2$ molecule is clearly detected, whose most prominent lines are the
transitions H$_2$ 1-0 (S2) and H$_2$ 1-0 (S1), although only H$_2$ 1-0 (S1) could
be measured. H$_2$ 1-0 (S3) is not clear in the 
nuclear spectra because it is too close to the [SiVI] line, which is very broad. 
The flux of H$_2$ can be used to derive the mass of hot molecular gas, as shown in section 4.3.

Several other lines of lower intensity are detected in the ZJ spectra, namely 
[CaI]$\lambda$0.985~$\micron$, [SVIII]$\lambda$0.991~$\micron$, 
HeII$\lambda$1.012~$\micron$, [SII]$\lambda$1.032~$\micron$ (a blend of 4 [SII] transitions),
[PII]$\lambda$1.188~$\micron$, [FeII]$\lambda$1.257~$\micron$ and the coronal 
line [SIX]$\lambda$1.252~$\micron$. Wherever possible, emission line fluxes for all features were 
measured assuming that the line profiles can be represented by a single or a sum of
gaussian profiles. The continuum underneath each line was fit by a low-order
polynomium, usually a straight line. The SMART routine (Higdon et al. 2004), a software 
developed by the Infrared Spectrograph (IRS) Instrument Team, 
originally designed to perform real-time processing and analysis of IRS data,
was used for this purpose. 

The results are shown in Table~1. Errors on the fluxes were obtained from the 
values given by
SMART, which is calculated based on the residual of the fit and the flux
calibration error, which is about 10 \% for the infrared.

The detection of [P\,{\sc ii}] with
considerable strength is particularly interesting. As explained by 
Mazzalay \& Rodr\'{\i}guez-Ardila (2007), in the Sun 
this element is a factor $\sim$1000 times less abundant than
carbon. In the nuclear spectrum of NGC~1068
we find a ratio of [P\,{\sc ii}]/[C\,{\sc i}] equal to 0.8. If the P/C abundance is
about solar, the [P\,{\sc ii}] line should not 
be present, unless carbon, like iron, might be locked in grains.
This indicates that phosphorus may be overabundant by
a factor of 10-20 compared to its solar abundance, relative to hydrogen.
 
The observed line widths are between 800 - 1000 km~s$^{-1}$. This is in agreement
with what is found in the optical region (Pelat \& Alloin 1980).
Although NGC 1068 has been considered a prototypical Seyfert 2 galaxy,
its forbidden lines are much wider than those of typical Seyfert 2s. Moreover,
some emission-line profiles show multiple components (Pelat \& Alloin 1980, Alloin
et al. 1983, Meaburn \& Pedlar 1986) also seen in the NIR as we show in section 4.1.

\begin{table}
\caption{Observed nuclear emission-line fluxes and EWs}
\begin {tabular} {@{}|l|ccc}
\hline
        &   $\lambda$   &    Flux                                & EW  \\
Line    &   ($\micron$) &   (10$^{-13}$ ergs cm$^{-2}$ s$^{-1}$) & ($\micron$)  \\
\hline
[S III]       & 0.907   &  7.50 $\pm$ 0.77 & 0.0085 $\pm$ 0.0008 \\
$[$S III$]$   & 0.953   & 19.80 $\pm$ 2.03 & 0.0242 $\pm$ 0.0026\\
$[$C I$]$     & 0.985   &  0.91 $\pm$ 0.11 & 0.0024 $\pm$ 0.0003\\
$[$S VIII$]$  & 0.991   &  0.91 $\pm$ 0.11 & 0.0015 $\pm$ 0.0002\\
He II         & 1.012   &  1.61 $\pm$ 0.17 & 0.0023 $\pm$ 0.0006\\
$[$S II$]$    & 1.032   &  2.81 $\pm$ 0.36 & 0.0035 $\pm$ 0.0008\\
He I          & 1.083   & 17.00 $\pm$ 1.76 & 0.0327 $\pm$ 0.0071\\
Pa$\gamma$    & 1.093   &  0.97 $\pm$ 0.12 & 0.0026 $\pm$ 0.0004\\
$[$PII$]$     & 1.188   &  0.78 $\pm$ 0.14 & 0.0010 $\pm$ 0.0002\\
$[$S IX$]$    & 1.252   &  0.88 $\pm$ 0.12 & 0.0021 $\pm$ 0.0003\\
$[$Fe II$]$   & 1.257   &  0.91 $\pm$ 0.13 & 0.0023 $\pm$ 0.0012\\
Pa$\beta$     & 1.282   &  2.24 $\pm$ 0.29 & 0.0033 $\pm$ 0.0007\\
$[$Si X$]$    & 1.430   &  1.59 $\pm$ 0.18 & 0.0021 $\pm$ 0.0007\\
$[$Fe II$]$   & 1.643   &  0.72 $\pm$ 0.09 & 0.0008 $\pm$ 0.0002\\
H$_2$ 1-0 S(3)& 1.956   &  0.54 $\pm$ 0.07 & 0.0003 $\pm$ 0.0001\\
$[$Si VI$]$   & 1.962   &  3.39 $\pm$ 0.37 & 0.0030 $\pm$ 0.0009\\
H$_2$ 1-0 S(1)& 2.121   &  0.35 $\pm$ 0.05 & 0.0002 $\pm$ 0.0001\\
Br$\gamma$    & 2.165   &  0.42 $\pm$ 0.06 & 0.0002 $\pm$ 0.0001\\
$[$Ca VIII$]$ & 2.322   &  1.01 $\pm$ 0.23 & 0.0005 $\pm$ 0.0002\\
\hline
\end{tabular}
\end{table}

\begin{figure*} 
\includegraphics [width=170mm]{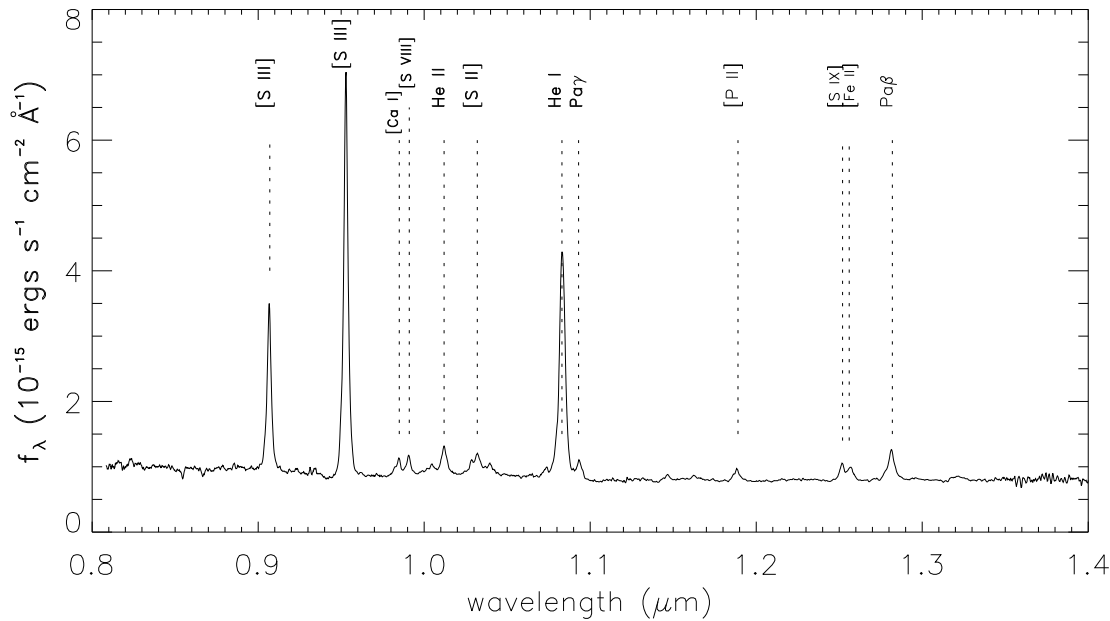}
\caption{Flux calibrated spectrum in the ZJ range of the 
nuclear region of NGC 1068. The aperture width is 1".6. Major emission lines detected
are identified.}
\end{figure*}

\begin{figure*} 
\includegraphics [width=170mm]{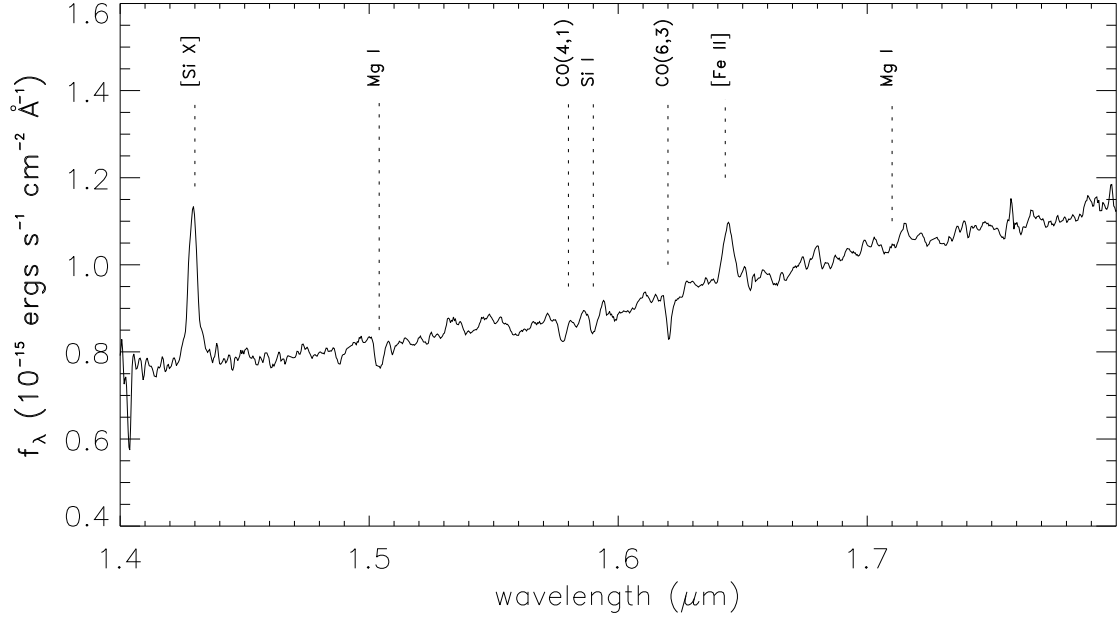}
\caption{Same as Figure 2, but in the H band. }
\end{figure*}

\begin{figure*} 
\includegraphics [width=170mm]{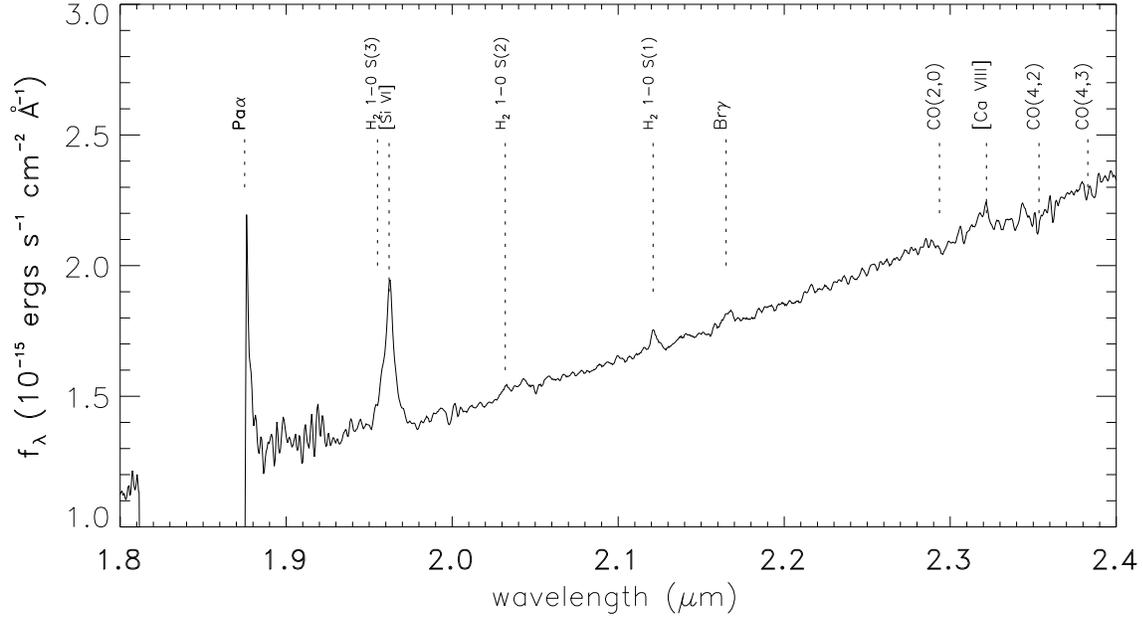}
\caption{Same as Figure 2, but in the K band. Note that P$\alpha$ is severely affected by telluric absorption.}
\end{figure*}

\subsection{Absorption features}

Several absorption features are clearly seen in the H- and K-band 
nuclear spectrum (Figures 3 and 4). In particular, we observe the 
following ones: MgI 1.50 $\micron$, CO 1.58 $\micron$, SiI 1.59 $\micron$, 
CO 1.62 $\micron$, MgI 1.71 $\micron$, CO 2.29 $\micron$, 
CO~2.35 $\micron$ and CO 2.38 $\micron$. These features
are identified after their main contributor, although,
depending on the dominant stellar type, other features
may become dominant.

The CO 1.62 $\micron$ can be used to obtain a rough estimate of the contribution
of the stellar population flux to the H band. This is done
by measuring the depth of the $^{12}$CO(6-3) overtone bandhead at 1.62 $\micron$.
We found that the observed depth is 12$\%$ $\pm$ 1$\%$
of the continuum. This can be compared to the $\sim$ 20$\%$
that is typically expected for a population of GKM supergiants that dominates the H 
band light for a stellar population older than 10$^7$ yr (Schinnerer et al. 1998).
It can thus be roughly estimated that about 60$\%$ $\pm$ 5$\%$
of the total flux in the H band continuum is due to GMK supergiants.
Following Schinnerer et al. (1998), although these
stars dominate the stellar emission in the H band, about 
one-third of the total stellar flux comes from
stars of other stellar classes. Therefore, 90$\%$ $\pm$ 8$\%$
of the nuclear H band continuum could be of stellar origin.
This is in reasonable agreement with the 74$\%$ $\pm$ 8$\%$ 
determined by Cid Fernandes et al. (2001) for this object, based 
on optical spectroscopy. This means that the H-band continuum
is essentially dominated by the stellar population, with only
$\sim$~10 - 20 $\%$ due to the AGN.
This calculations show how the NIR spectroscopy is a potential
tool to unveil hidden starburst components, either in highly
obscured objects or in those in which the optical continuum emission is
dominated by the AGN component (as in Seyfert 1 galaxies).  

One way of estimating the age of the stellar population in the
NIR is using the CO index defined by Ivanov et al. (2000),
for the absorption feature 
starting at 2.29 $\micron$. This index is supposed to be insensitive to extinction 
and to possible uncertainties in the continuum shape of the infrared
spectra and it is defined as

\begin{equation}
  \label{equ:coind}
  CO= - 2.5 log \left( \frac{\langle F_{2.295 }\rangle}{\langle F_{2.282}\rangle}\right)
\end{equation}

where $\langle F_{2.282} \rangle$ and $\langle F_{2.295} \rangle$ are the average
fluxes within a bandwidth of 0.01 $\micron$, centered on the blue continuum and on the
bandhead respectively. For the nuclear aperture we obtained an index
of -0.01 $\pm$ 0.05. This value is extremely
low for a Seyfert 2 galaxy, according to the sample of Ivanov et al. 
Notice that in a Seyfert galaxy nucleus, the
late-type stellar features are substantially diluted by non-stellar nuclear
emission. NGC~1068 is probably an extreme case, since its nuclear spectrum
shows a very steep increase in the K band continuum flux with wavelength 
(Figure 4), much steeper than what is expected for a non-thermal power-law source. 
The K band excess of emission is
probably due to thermal radiation from hot dust (Thatte et al. 1997).

In order to minimize the influence of the shape of the continuum, we normalized
the spectrum in this region using a linear fit in the region between
2.20 and 2.26 $\micron$, where there are no apparent features. After dividing the spectrum
by this fit the CO index is recalculated. The new value is 0.01 $\pm$ 0.05. 
Thus, the low value is not caused by the shape of the spectrum and  
suggests that there is no evidence for a strong starburst, in agreement with the
results of Ivanov et al. (2000) for Seyfert galaxies, but contrary to what was found
in the H band. 

Nevertheless, it is really hard to estimate the true dilution of the bandhead by the continuum. 
The best way would be to do a full stellar population fitting, including
a non-stellar component. This is out of the scope of this paper, 
but is presented in detail in a follow-up paper (Martins et al. 2010).

\section{Spectra of the Extended Region}

In addition to the nuclear spectra discussed in the preceding section,
the spatial information contained in the slit allowed us to extract the
spectra corresponding to the narrow line region and to the extended line
region of NGC~1068 to the limits of $\sim$350~pc north and south of 
the nucleus. Gas kinematics as well as extinction and the
molecular and [Fe\,{\sc ii}] lines will be studied in more detail
in the following sections. 

\subsection{Gas kinematics}
The spectra for all integrated apertures are presented in Figures 5-7, where 
the wavelengths are in the observer's rest frame.
It is easy to see that the most conspicuous lines show a double peaked profile. 
One component is in the rest frame of the galaxy (primary component, hereafter) 
while the other (second component, hereafter) is either blueshifted towards the north
of the nucleus or redshifted to the south. The relative intensity of the peaks also 
varies with distance, as shown in Figure~8 for [S\,{\sc iii}]$\lambda$0.953~$\micron$, the
strongest line in our spectra. It can be seen that [S\,{\sc iii}]
is detected at distances of up to $\sim$350 pc from the nucleus.
As explained in Sec.~2, for the purpose of comparing the line profiles 
we divided the nuclear spectra (1.6'' aperture) into seven smaller apertures (see Fig.~1). 
The aperture named ``nuc$\_a$'',  centered on the peak of the continuum light
distribution, is extracted over the central 0.4" window and the remaining six with 
apertures 0.2" wide. Although the integrated flux of these smaller apertures
are oversampled because the extraction window is smaller than the seeing value,
this approach allows us to study line profiles, which clearly
show changes along the spatial direction at scales down to 20~pc.
 
A comparison of the line profiles in velocity space observed in the most 
conspicuous emission lines is presented in Figure 9. In this figure, the continuum underneath
each line was adjusted using a linear interpolation and then subtracted. It can be seen that the coronal line
[S\,{\sc viii}] shows a profile very similar to that of [S\,{\sc iii}], and second components
can be clearly seen. The molecular H$_2$ lines also 
show scarce evidence of a second component, not so clear since the line is weak.
They are also narrower than the emission 
features of other species outside the nucleus, which indicates that it might be formed in a different 
region. Indeed, NIR observations has shown that the H$_2$ emission is distributed
mainly along a direction perpendicular to the ionizing cone (Galliano \& Alloin 2002). 
The [Fe\,{\sc ii}] lines seem to follow the H$_2$ lines more closely.

\begin{figure*} 
\includegraphics [width=170mm]{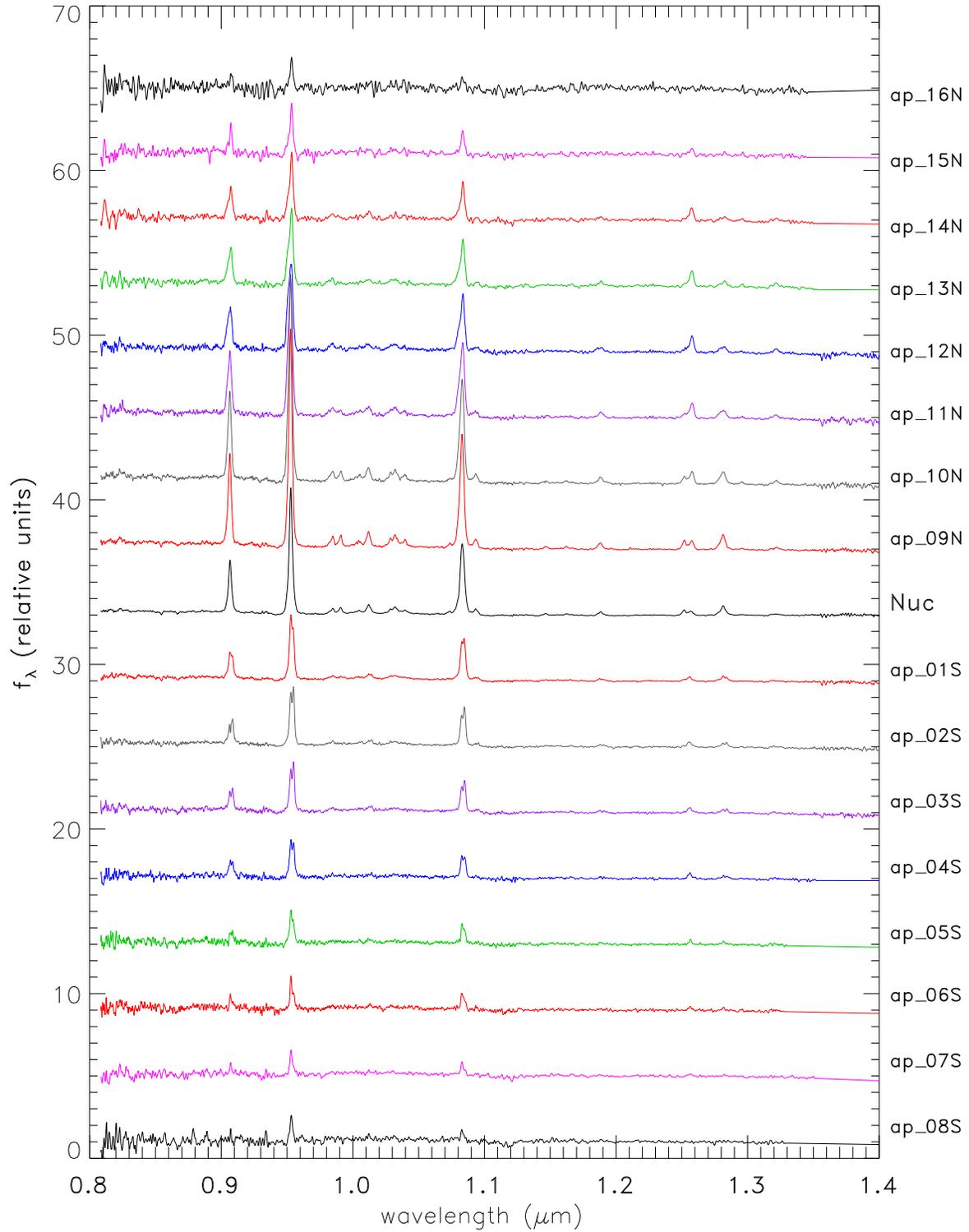}
\caption{Spectra of all apertures of NGC 1068 in the ZJ range. From the top to the bottom
the spectra correspond to the apertures from north to south. }

\end{figure*}

\begin{figure*} 
\includegraphics [width=170mm]{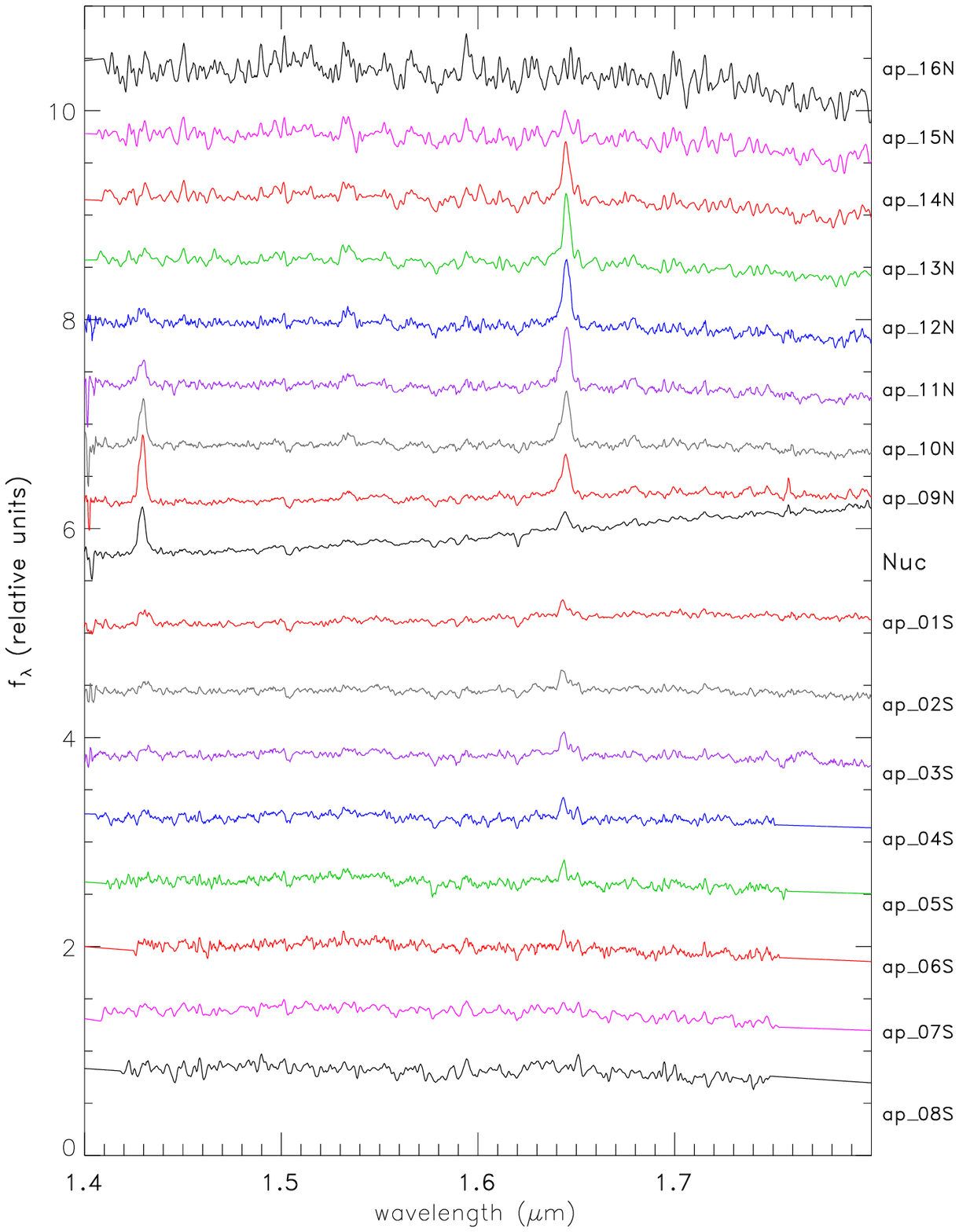}
\caption{Same as in Figure 5, but in the H band.}
\end{figure*}

\begin{figure*} 
\includegraphics [width=170mm]{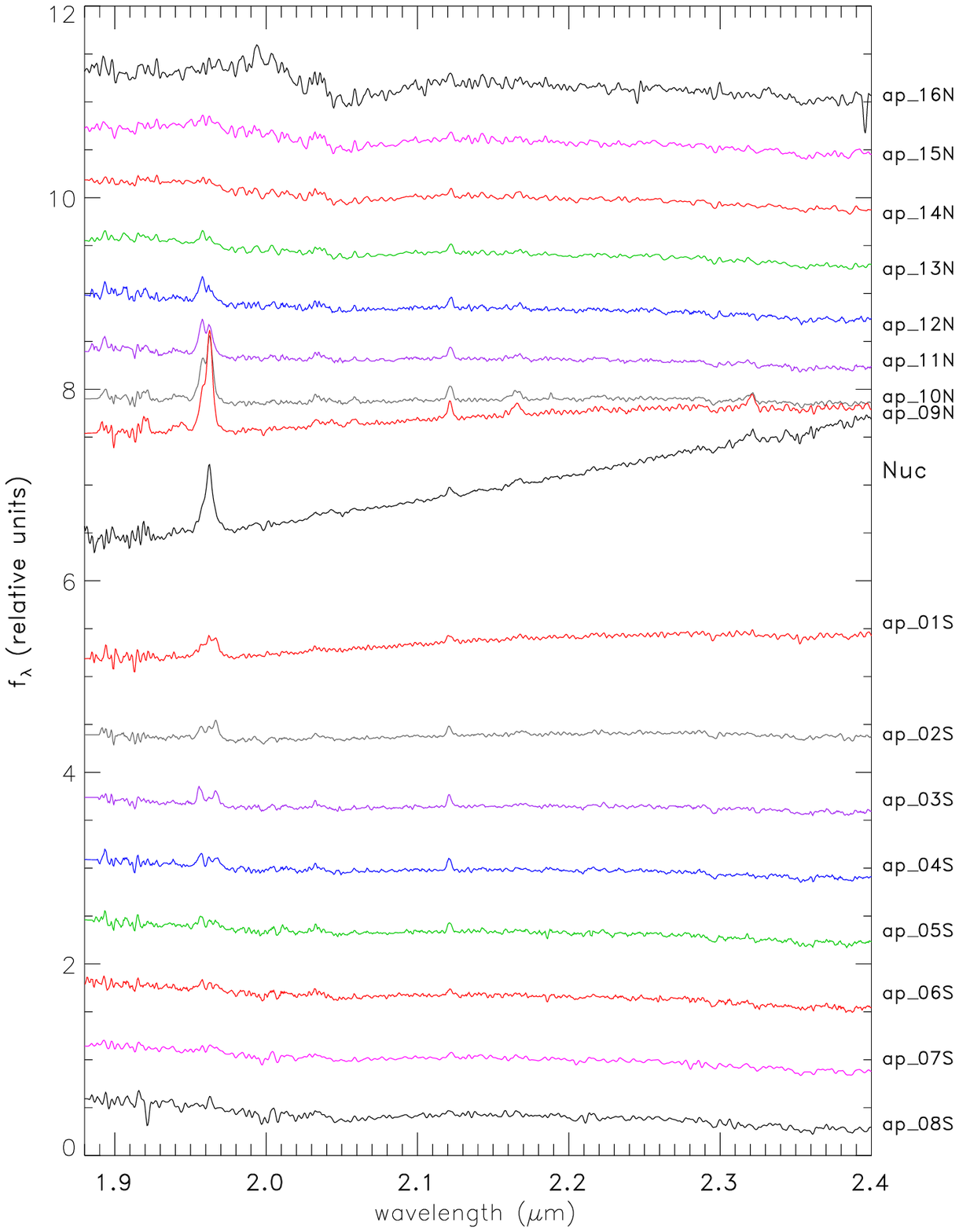}
\caption{Same as in Figure 5 but in the K band.}
\end{figure*}

\begin{figure*}
\label{nirzoom} 
\includegraphics [width=150mm]{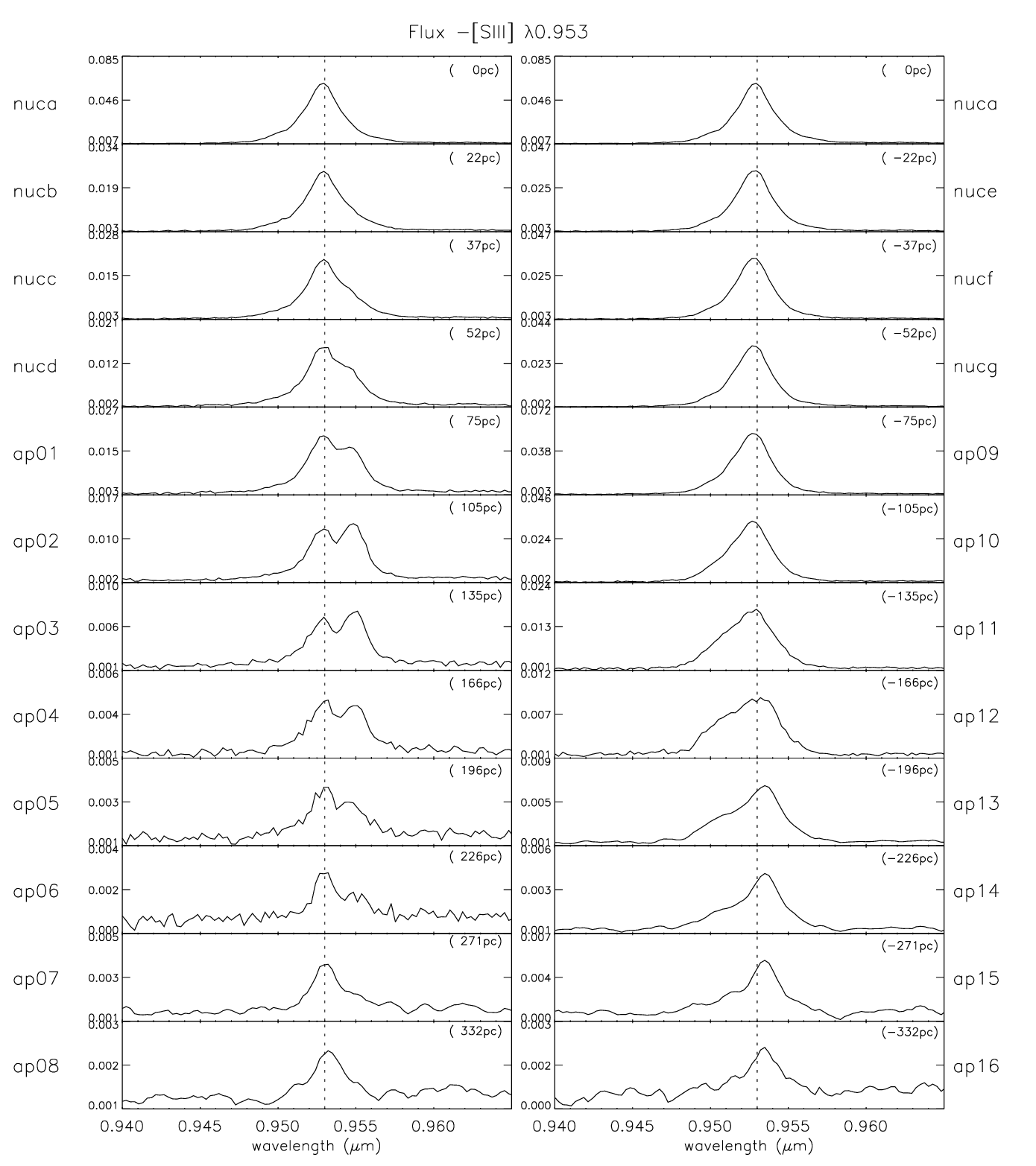}
\caption{Profile of the [SIII]$\lambda$0.953 $\micron$ for all apertures.
Left panels show the apertures to the south, where the second peak is
redshifted. Right panels are the apertures to the north, where the second peak
is blueshifted. The first plot in each column is the very central
aperture, plotted in both columns for comparison.}
\end{figure*}

We did not identify a broad component with FWHM $\sim$~2000~km~s$^{-1}$ in any line,
although they have been
reported on FUSE observations (Zheng et al. 2008). That may be because FUSE
slit collects emission from a block of regions spanning several arcseconds perpendicular
to the conical axis and the total line emission from these regions may be blended
into one broad component. Since we are observing much smaller regions with each 
aperture, we minimize this effect.

Not all emission lines have double peaked profiles. This is the case of 
high ionization lines such as [S\,{\sc viii}] and [Si\,{\sc x}] as well as molecular
lines of H$_2$. Moreover, low ionization lines like [Ca\,{\sc i}] or [P\,{\sc ii}] also
seem to be devoided of a double-peak structure but these 
are faint lines. The lack of splitting may be either by intrinsic reasons or due
to low S/N. In Figure~10, the relative flux of the most conspicuous
emission lines are shown as a function of the distance to the center. The fluxes
were normalized relative to  the nucleus to allow the comparison between the
different lines. Negative distances represent the northern direction and positive 
distances the southern one. For double peaked lines, the upper panel shows the
distribution of the primary component while the bottom panel that of the
secondary component.
It can be seen that the second component of [S\,{\sc iii}]~0.953$\mu$m and HeI~1.083$\mu$m
are observed up to 200~pc from the nucleus. Indeed, it seems that
there is a maximum of the second components between 100 - 150~pc from the center.
As for the primary components, evidence of emission up to 300~pc from the center
is observed although it becomes too much faint to be measured at 3$\sigma$ level. 
Note that the [Fe\,{\sc ii}] emission is the most extended of all lines analyzed.

The flux gas distribution shown in Figure~10 and mapped by the first time in the
literature with data in the interval 0.8$-2.4\mu$m confirms previous reports in the 
optical and MIR in the sense that the northern zone has stronger emission than the 
southern one (Mazzalay et al. 2009; Geballe et al. 2009). This can be attributed 
to the strong extinction towards the observer. In this scenario, the ionization cone 
is partially hidden from our view because it crosses the host galaxy plane.
Note also that high ionization lines like 
[Si\,{\sc x}]~1.430$\mu$m and [Si\,{\sc vi}]~1.954$\mu$m are detected only up to
$\sim$200~pc from the nucleus. This is a clear indication that the gas is more ionized near
the nucleus.

The observed emission features of the extended region were also measured with the
software SMART. The results are shown in Tables 2 and 3. Errors were estimated
the same way as in Table 1.

\begin{table*}
\caption{Observed emission-line fluxes of all South non-nuclear apertures}

\begin {tabular} {@{}|l||c cccccccc}
\hline \hline
 \multicolumn{8}{|c|}
 {\rule[-0mm]{0mm}{3mm} South - Main components}\\
\hline \hline
 & & \multicolumn{6}{|l|}
{\rule[-0mm]{0mm}{3mm} Flux (10$^{-14}$ ergs cm$^{-2}$ s$^{-1}$)}\\
\hline
Line    & $\lambda$ ($\micron$)   &  Ap01   & Ap02         & Ap03            & Ap04            & Ap05            & Ap06            & Ap07    & Ap08  \\ 
  \hline
$[$S III$]$   & 0.907 &  5.81 $\pm$ 1.10 & 1.63 $\pm$ 0.27 & 1.28 $\pm$ 0.25 & 1.31 $\pm$ 0.22 & 1.04 $\pm$ 0.19 & 0.45 $\pm$ 0.08 & 0.61 $\pm$ 0.06 & 0.23 $\pm$ 0.03 \\
$[$S III$]$   & 0.953 & 14.21 $\pm$ 1.80 & 6.66 $\pm$ 0.85 & 3.95 $\pm$ 0.71 & 3.93 $\pm$ 0.70 & 1.84 $\pm$ 0.33 & 0.96 $\pm$ 0.14 & 1.90 $\pm$ 0.24 & 1.22 $\pm$ 0.15 \\
$[$C I$]$     & 0.985 &  1.60 $\pm$ 0.26 & 0.48 $\pm$ 0.09 & 0.45 $\pm$ 0.11 & 0.64 $\pm$ 0.13 &  -              & -               & -               &  -              \\
$[$S VIII$]$  & 0.991 &  0.21 $\pm$ 0.07 & -               &  -              & -               &  -              & -               & -               &  -              \\
He II         & 1.012 &  1.12 $\pm$ 0.37 & 1.00 $\pm$ 0.15 & 0.63 $\pm$ 0.16 & 0.61 $\pm$ 0.12 & 0.34 $\pm$ 0.10 & -               & 0.11 $\pm$ 0.02 & 0.22 $\pm$ 0.03 \\
$[$S II$]$    & 1.032 &  3.50 $\pm$ 0.77 & 1.28 $\pm$ 0.28 & 0.81 $\pm$ 0.16 & -               & -               & -               & 0.19 $\pm$ 0.03 &  -              \\
He I          & 1.083 & 11.48 $\pm$ 1.56 & 3.64 $\pm$ 0.48 & 2.36 $\pm$ 0.32 & 2.38 $\pm$ 0.43 & 1.43 $\pm$ 0.34 & 0.77 $\pm$ 0.11 & 0.84 $\pm$ 0.12 & 0.58 $\pm$ 0.08 \\
Pa$\gamma$    & 1.093 &  1.13 $\pm$ 0.19 & 0.74 $\pm$ 1.45 & 0.58 $\pm$ 0.13 & 0.62 $\pm$ 0.14 & -               & 0.07 $\pm$ 0.01 & 0.08 $\pm$ 0.01 &  -              \\
$[$PII$]$     & 1.188 &  1.10 $\pm$ 0.19 & 0.82 $\pm$ 0.12 & 0.64 $\pm$ 0.10 & 1.01 $\pm$ 0.13 & -               & -               & -               & -               \\
$[$S IX$]$    & 1.252 &  0.39 $\pm$ 0.08 & 0.11 $\pm$ 0.02 & 0.72 $\pm$ 0.10 & 0.22 $\pm$ 0.09 & -               & -               & -               &  -              \\
$[$Fe II$]$   & 1.257 &  1.39 $\pm$ 0.28 & 1.06 $\pm$ 0.15 &  -              & 0.37 $\pm$ 0.08 & 0.28 $\pm$ 0.07 & 0.22 $\pm$ 0.05 & 0.36 $\pm$ 0.05 &  -              \\
Pa$\beta$     & 1.282 &  2.22 $\pm$ 0.41 & 1.04 $\pm$ 0.14 & 1.06 $\pm$ 0.15 & 0.65 $\pm$ 0.12 & 0.39 $\pm$ 0.09 & 0.13 $\pm$ 0.03 & 0.12 $\pm$ 0.02 &  -              \\
$[$Si X$]$    & 1.430 &  1.43 $\pm$ 0.27 & 1.25 $\pm$ 1.05 & -               & -               & -               & -               & -               &  -              \\
$[$Fe II$]$   & 1.643 &  0.76 $\pm$ 0.59 & 1.27 $\pm$ 0.17 & 0.57 $\pm$ 0.07 & 0.58 $\pm$ 0.08 & 0.33 $\pm$ 0.06 & 0.15 $\pm$ 0.03 & -               &  -              \\
$[$Si VI$]$   & 1.962 &  3.28 $\pm$ 0.41 & 1.06 $\pm$ 0.17 & 0.17 $\pm$ 0.03 & 0.16 $\pm$ 0.03 & -               & -               & 0.23 $\pm$ 0.04 & 0.12  $\pm$ 0.01 \\
H$_2$ 1-0 S(1)& 2.121 &  0.65 $\pm$ 0.13 & 0.32 $\pm$ 0.04 & 0.24 $\pm$ 0.03 & 0.25 $\pm$ 0.03 & 0.14 $\pm$ 0.02 & 0.10 $\pm$ 0.02 & 0.12 $\pm$ 0.02 & 0.10  $\pm$ 0.01 \\
Br$\gamma$    & 2.165 &   -              & -               & -               & -               & -               &                 & -               &  -               \\
$[$Ca VIII$]$ & 2.322 &  0.21 $\pm$ 0.03 & -               & -               & -               & -               &                 & -               &  -               \\
\hline
\hline
  \multicolumn{8}{|c|}
 {\rule[-0mm]{0mm}{3mm} South - Second components}\\
\hline \hline
 & & \multicolumn{6}{|l|}
{\rule[-0mm]{0mm}{3mm} Flux (10$^{-14}$ ergs cm$^{-2}$ s$^{-1}$)}\\
\hline
Line  & $\lambda$ ($\micron$)&  Ap01    & Ap02            & Ap03            & Ap04             & Ap05           & Ap06          & Ap07 & Ap08  \\ 
  \hline
$[$S III$]$   & 0.907 & 2.13 $\pm$ 0.44 & 2.77 $\pm$ 0.52 & 1.25 $\pm$ 0.13 & 1.24 $\pm$ 0.24 & -               & -               & -     & -\\
$[$S III$]$   & 0.953 & 4.16 $\pm$ 0.81 & 4.81 $\pm$ 0.87 & 2.87 $\pm$ 0.31 & 2.87 $\pm$ 0.54 & 0.76 $\pm$ 0.16 & 0.65 $\pm$ 0.14 & -     &- \\
He II         & 1.012 & 0.17 $\pm$ 0.31 & -               &   -             & -               & -               & -               & -      &   - \\
He I          & 1.083 & 4.09 $\pm$ 0.79 & 4.46 $\pm$ 0.82 & 2.20 $\pm$ 0.56 & 2.18 $\pm$ 0.43 & 0.69 $\pm$ 0.14 & 0.29 $\pm$ 0.07 & 0.25 $\pm$ 0.07 &-   \\
Pa$\beta$     & 1.282 & 0.20 $\pm$ 0.04 & 0.29 $\pm$ 0.06 & 0.18 $\pm$ 0.09 & 0.76 $\pm$ 0.16 & -               & -               &  -     &-   \\
$[$Si VI$]$   & 1.962 &  -              & 0.76 $\pm$ 0.14 & 0.35 $\pm$ 0.11 & 0.39 $\pm$ 0.09 & -               & -               &  -     &-   \\

\hline
\end{tabular}
\end{table*}

\begin{table*}
\caption{Observed emission-line fluxes of all North non-nuclear apertures}

\begin {tabular} {@{}|l||c cccccccc}
\hline
\hline
 \multicolumn{8}{|c|}
 {\rule[-0mm]{0mm}{3mm} North - Main components}\\
\hline \hline
 & & \multicolumn{6}{|l|}
{\rule[-0mm]{0mm}{3mm} Flux (10$^{-14}$ ergs cm$^{-2}$ s$^{-1}$)}\\
\hline
Line   & $\lambda$($\micron$)& Ap09      & Ap10             & Ap11             & Ap12            & Ap13            & Ap14            & Ap15 & Ap16  \\ 
  \hline
$[$S III$]$   & 0.907 & 17.96 $\pm$ 2.06 & 11.73 $\pm$ 1.18 &  6.50 $\pm$ 0.96 & 2.53 $\pm$ 0.33 & 1.74 $\pm$ 0.24 & 1.43 $\pm$ 0.19 & 1.32 $\pm$ 0.18 & 0.56 $\pm$ 0.09 \\
$[$S III$]$   & 0.953 & 46.14 $\pm$ 6.96 & 30.03 $\pm$ 3.01 & 16.33 $\pm$ 2.10 & 7.46 $\pm$ 0.99 & 4.83 $\pm$ 0.62 & 2.94 $\pm$ 0.39 & 3.25 $\pm$ 0.45 & 1.24 $\pm$ 0.15   \\
$[$C I$]$     & 0.985 &  2.38 $\pm$ 0.42 &  1.85 $\pm$ 0.27 &  1.39 $\pm$ 0.20 & 0.88 $\pm$ 0.13 & 0.53 $\pm$ 0.08 & 0.55 $\pm$ 0.07 & -               & -    \\
$[$S VIII$]$  & 0.991 &  1.86 $\pm$ 0.30 &  1.40 $\pm$ 0.21 &  0.67 $\pm$ 0.10 & 0.21 $\pm$ 0.04 & -               & -               & -               & -   \\
He II         & 1.012 &  4.25 $\pm$ 0.70 &  2.19 $\pm$ 0.30 &  1.09 $\pm$ 0.16 & 0.68 $\pm$ 0.10 & 0.50 $\pm$ 0.08 & 0.57 $\pm$ 0.08 & -               & -   \\
$[$S II$]$    & 1.032 &  6.36 $\pm$ 1.05 &  4.15 $\pm$ 0.57 &  2.30 $\pm$ 0.30 & 0.90 $\pm$ 0.12 & 0.69 $\pm$ 0.09 & 0.30 $\pm$ 0.05 & -               & -  \\
He I          & 1.083 & 29.19 $\pm$ 4.34 & 17.39 $\pm$ 1.75 &  9.87 $\pm$ 1.33 & 4.75 $\pm$ 0.65 & 2.31 $\pm$ 0.30 & 2.19 $\pm$ 0.30 & 1.98 $\pm$ 0.22 & 0.74 $\pm$ 0.13 \\
Pa$\gamma$    & 1.093 &  2.46 $\pm$ 0.31 &  1.45 $\pm$ 0.20 &  1.10 $\pm$ 0.19 & 0.72 $\pm$ 0.13 & 0.30 $\pm$ 0.05 & 0.38 $\pm$ 0.05 & -               & -  \\
$[$PII$]$     & 1.188 &  2.01 $\pm$ 0.22 &  1.46 $\pm$ 0.19 &  1.01 $\pm$ 0.13 & 1.01 $\pm$ 0.13 & 0.39 $\pm$ 0.06 & -               & -               & -  \\
$[$S IX$]$    & 1.252 &  2.31 $\pm$ 0.28 &  1.68 $\pm$ 0.23 &  0.81 $\pm$ 0.12 & 0.99 $\pm$ 0.16 & 0.36 $\pm$ 0.06 & 0.22 $\pm$ 0.04 & 0.15 $\pm$ 0.03 & -  \\
$[$Fe II$]$   & 1.257 &  2.40 $\pm$ 0.29 &  2.06 $\pm$ 0.28 &  1.95 $\pm$ 0.27 & 1.14 $\pm$ 0.19 & 1.15 $\pm$ 0.17 & 0.77 $\pm$ 0.10 & 0.51 $\pm$ 0.07 & -  \\
Pa$\beta$     & 1.282 &  4.81 $\pm$ 0.70 &  2.44 $\pm$ 0.32 &  1.02 $\pm$ 0.42 & 0.85 $\pm$ 0.12 & 0.52 $\pm$ 0.08 & 0.27 $\pm$ 0.04 & 0.24 $\pm$ 0.04 & -  \\
$[$Si X$]$    & 1.430 &  2.41 $\pm$ 0.30 &  1.06 $\pm$ 0.15 &  0.76 $\pm$ 0.11 & 0.45 $\pm$ 0.09 & -               & -               &  -              & -   \\
$[$Fe II$]$   & 1.643 &  1.91 $\pm$ 0.26 &  1.71 $\pm$ 0.25 &  1.80 $\pm$ 0.21 & 1.43 $\pm$ 0.16 & 1.00 $\pm$ 0.12 & 0.79 $\pm$ 0.09 & 0.78 $\pm$ 0.10 & -  \\
$[$Si VI$]$   & 1.962 &  5.51 $\pm$ 0.69 &  2.42 $\pm$ 0.36 &  0.86 $\pm$ 0.12 & 0.43 $\pm$ 0.10 & 0.13 $\pm$ 0.03 & -               & -               & -  \\
H$_2$ 1-0 S(1)& 2.121 &  0.76 $\pm$ 0.10 &  0.53 $\pm$ 0.06 &  0.32 $\pm$ 0.04 & 0.21 $\pm$ 0.04 & 0.13 $\pm$ 0.02 & 0.08 $\pm$ 0.01 & 0.15 $\pm$ 0.03 & -   \\
Br$\gamma$    & 2.165 &  0.99 $\pm$ 0.12 &  0.58 $\pm$ 0.09 &  -               & 0.16 $\pm$ 0.04 & -               & 0.12 $\pm$ 0.01 & -               & -\\
$[$Ca VIII$]$ & 2.322 &  1.14 $\pm$ 0.17 &  0.39 $\pm$ 0.09 &  -               & -               & -               & -               & -               & -  \\
\hline
\hline
 \multicolumn{8}{|c|}
 {\rule[-0mm]{0mm}{3mm} North - Second components}\\
\hline \hline
 & & \multicolumn{6}{|l|}
{\rule[-0mm]{0mm}{3mm} Flux (10$^{-14}$ ergs cm$^{-2}$ s$^{-1}$)}\\
\hline
Line  & $\lambda$ ($\micron$)&  Ap09    & Ap10            & Ap11            & Ap12            & Ap13            & Ap14            & Ap15       & Ap16  \\ 
  \hline
$[$S III$]$   & 0.907 & 2.49 $\pm$ 0.33 & 1.66 $\pm$ 0.22 & 1.20 $\pm$ 0.16 & 2.29 $\pm$ 0.33 & 1.18 $\pm$ 0.18 & 0.55 $\pm$ 0.08 & 0.43 $\pm$ 0.07 & -\\
$[$S III$]$   & 0.953 & 5.01 $\pm$ 0.70 & 3.90 $\pm$ 0.52 & 3.07 $\pm$ 0.45 & 3.33 $\pm$ 0.44 & 1.89 $\pm$ 0.28 & 1.20 $\pm$ 0.18 & 0.88 $\pm$ 0.13 & - \\
$[$S VIII$]$  & 0.991 & 0.47 $\pm$ 0.07 & -               & -               & -               & -               & -               & -          & - \\
He II         & 1.012 & -               & -               & -               & -               & -               & -               & -          & - \\
He I          & 1.083 & 4.23 $\pm$ 0.58 & 3.64 $\pm$ 0.49 & 1.98 $\pm$ 0.27 & 1.99 $\pm$ 0.29 & 2.07 $\pm$ 0.23 & 0.58 $\pm$ 0.09 & -          & -   \\
Pa$\beta$     & 1.282 & 0.58 $\pm$ 0.09 & 0.53 $\pm$ 0.08 & 0.62 $\pm$ 0.09 & -               & 0.19 $\pm$ 0.31 & 0.15 $\pm$ 0.03 & -          & -   \\
$[$Si X$]$    & 1.430 & 0.34 $\pm$ 0.05 & 0.25 $\pm$ 0.04 & -               & -               & -               & -               & -          & -   \\
$[$Fe II$]$   & 1.643 & 0.47 $\pm$ 0.06 & 0.46 $\pm$ 0.07 & -               & -               & -               & -               & -          & -  \\
$[$Si VI$]$   & 1.962 & 2.69 $\pm$ 0.35 & -               & -               & -               & -               & -               & -          &-   \\

\hline
\end{tabular}
\end{table*}

\begin{figure*} 
\includegraphics [width=170mm]{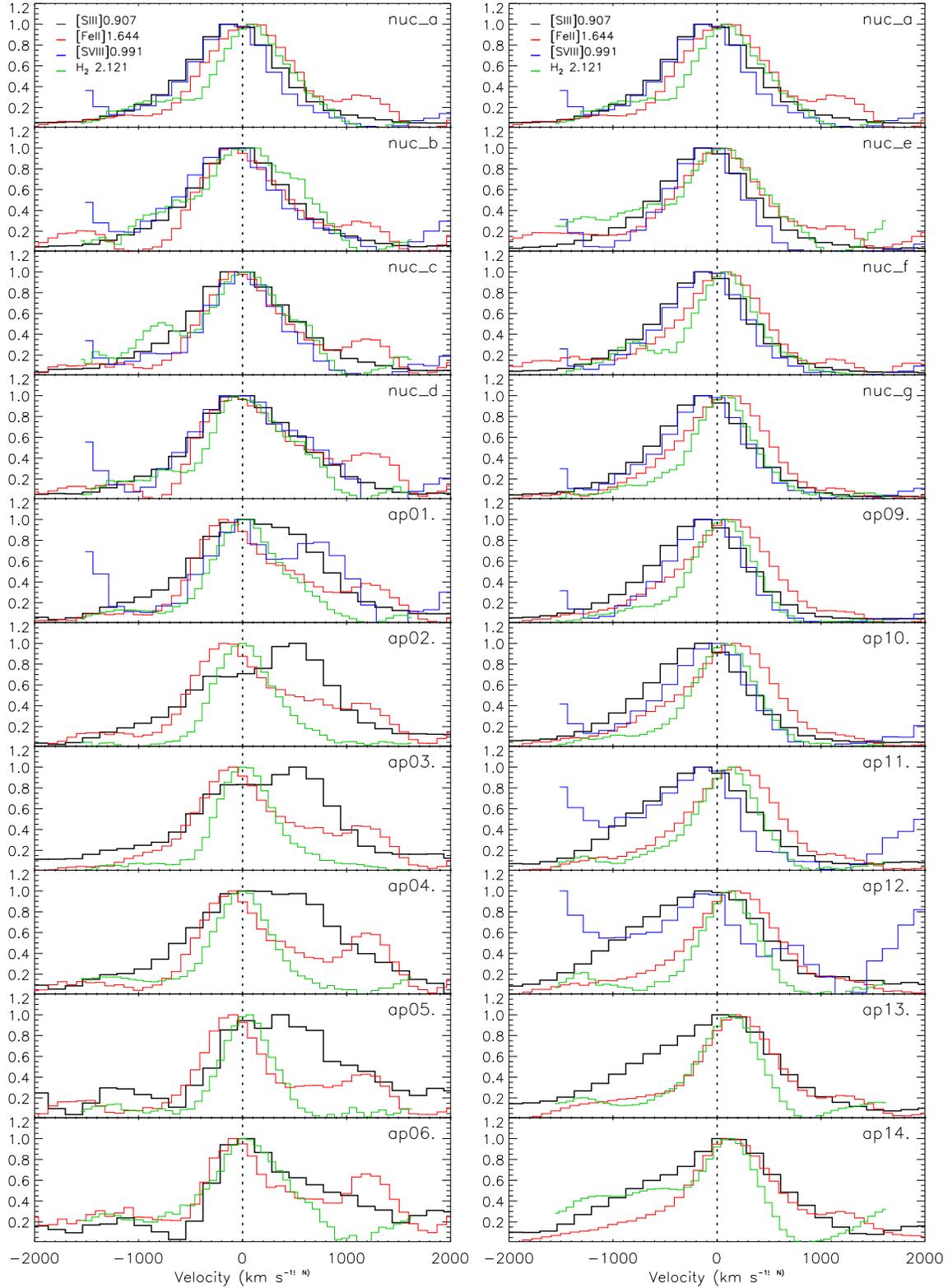}
\caption{Profile of emission lines in the velocity space. Left panels show the 
apertures to the south, where the second peak is redshifted. Right panels show
the apertures to the north, where the second peak is blueshifted. The first plot
in each column is the very central aperture, plotted in both columns for comparison. 
 Lines were only plotted for the apertures where they have measurable fluxes. }
\end{figure*}

\begin{figure*} 
\includegraphics [width=170mm]{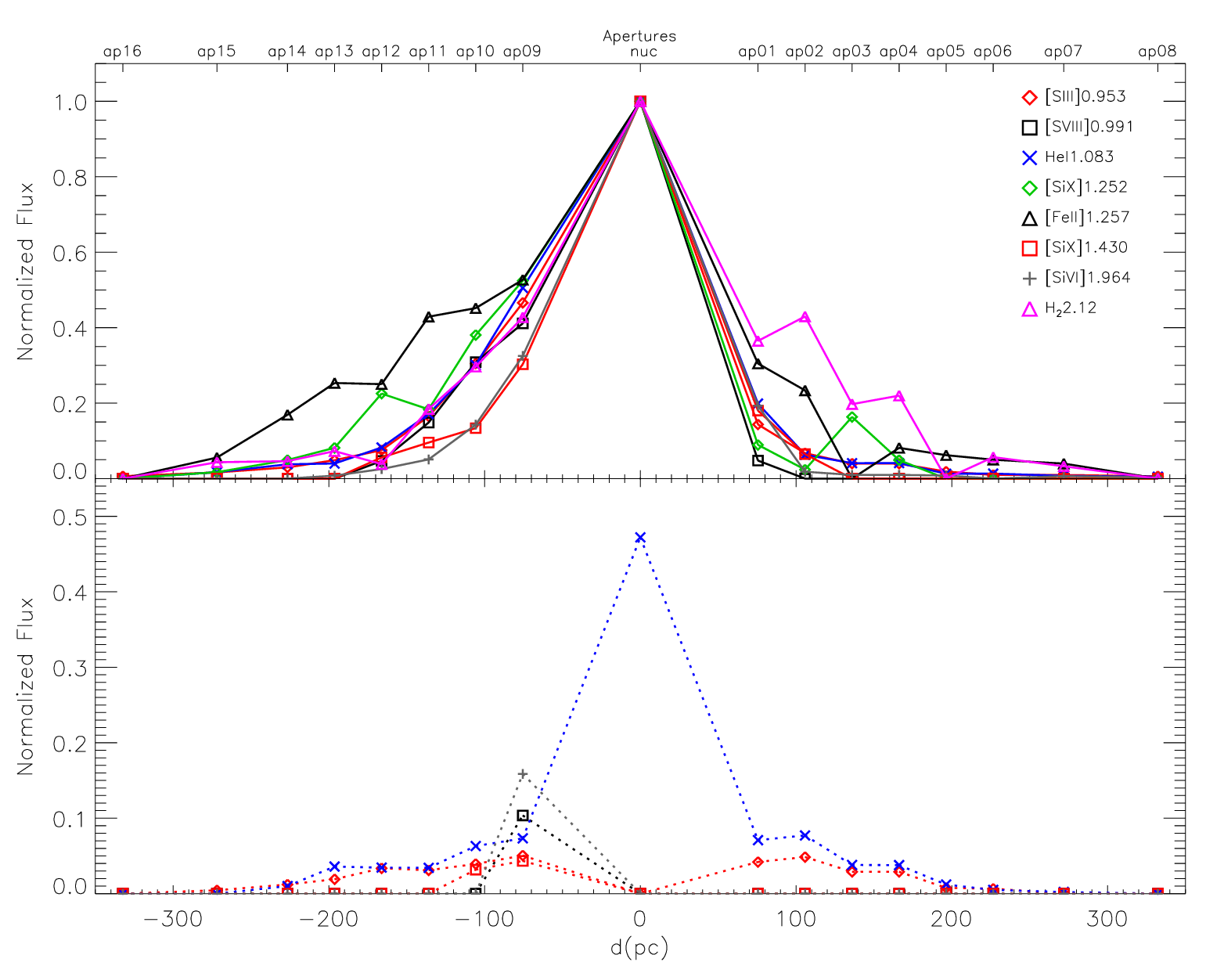}
\caption{Relative flux of main emission lines as a function of the distance to the center.
The fluxes were normalized by the center values to allow comparison between lines. Negative distances
represent the north direction and positive distances the south. The top panel shows the 
main components, while the bottom panel shows the second component if present.  }
\end{figure*}

\subsection{Extinction}

The extinction towards the nucleus and circumnuclear region was determined from 
the comparison of the predicted and observed emission
line ratios assuming the standard extinction 
law of Cardelli et al. (1989) for 
R$_V$ = 3.1. From our observed lines, two line ratios are extinction indicators: 
Pa$\beta$/Pa$\gamma$ and [Fe\,{\sc ii}]~1.257$\mu$m/1.643$\mu$m. 
We added both components of Pa$\beta$ for this purpose. The reason for this is
that Pa$\gamma$ has no second components, probably because it is a weak line and it was not
possible to distinguish them. When using only the primary component of Pa$\beta$, many ratios
give unphysical results.
Pa$\alpha$ could not be used in this calculation because it was
severely affected by telluric transmission. Moreover, Br$\gamma$ was only
detected in the central aperture. 

For hydrogen we used the intrinsic ratios for the 
case~B given by Osterbrock (1989). The pair of [Fe\,{\sc ii}] lines used to estimate 
the extinction share the same upper level, meaning that the intensity ratio depends only
on the energy differences between the lines and their Einstein-A coefficients, making
this line ratio a useful probe of reddening. 
The intrinsic value of the [Fe\,{\sc ii}]~1.257$\mu$m/1.643$\mu$m ratio was taken from
Bautista \& Pradhan (1998), calculated from
Nussbaumer \& Storey (1988) transition
probabilities. Table~4 shows the values of extinction found for each aperture. 
For the errors in this table we only considered the errors in the line measurements and
not in the flux calibration.

\begin{table}
\caption{Flux ratios and E(B-V)}
\begin {tabular} {@{}ccccc}
\hline
Line & Pa$\beta$/Pa$\gamma$  &   E(B-V)   &  [FeII]1.257/1.643 & E(B-V)  \\
\hline
Ap07 & 1.49 $\pm$ 0.31 & $<$ 0.01          & -               & -               \\ 
Ap06 & 1.84 $\pm$ 0.48 & 0.13 $\pm$ 0.25   & 1.54 $\pm$ 0.49 & $<$ 0.92        \\
Ap05 &        -        &    -              & 0.85 $\pm$ 0.22 & 1.65 $\pm$ 0.26 \\
Ap04 & 2.27 $\pm$ 0.55 & 1.05 $\pm$ 0.26   & 0.64 $\pm$ 0.17 & 2.65 $\pm$ 0.26 \\
Ap03 & 1.31 $\pm$ 0.55 & 0.82 $\pm$ 0.56   & 1.26 $\pm$ 0.16 & 0.28 $\pm$ 0.13 \\
Ap02 & 1.83 $\pm$ 0.41 & 0.09 $\pm$ 0.22   & 1.12 $\pm$ 0.31 & 0.69 $\pm$ 0.28 \\
Ap01 & 1.98 $\pm$ 0.51 & 0.44 $\pm$ 0.17   & 1.83 $\pm$ 0.51 & $<$ 0.10        \\
Nuc  & 2.31 $\pm$ 0.24 & 1.13 $\pm$ 0.10   & 1.26 $\pm$ 0.15 & 0.26 $\pm$ 0.12 \\
Ap09 & 1.96 $\pm$ 0.32 & 0.39 $\pm$ 0.12   & 1.26 $\pm$ 0.16 & 0.28 $\pm$ 0.13 \\
Ap10 & 2.05 $\pm$ 0.34 & 0.60 $\pm$ 0.13   & 1.20 $\pm$ 0.23 & 0.44 $\pm$ 0.20 \\
Ap11 & 1.50 $\pm$ 0.38 & $<$ 0.03          & 1.09 $\pm$ 0.10 & 0.80 $\pm$ 0.10 \\
Ap12 & 1.17 $\pm$ 0.35 & -                 & 0.80 $\pm$ 0.11 & 1.90 $\pm$ 0.14 \\
Ap13 & 2.33 $\pm$ 0.49 & 1.18 $\pm$ 0.18   & 1.15 $\pm$ 0.16 & 0.60 $\pm$ 0.14 \\
Ap14 & 1.11 $\pm$ 0.55 & $<$ 0.05          & 0.98 $\pm$ 0.26 & 1.17 $\pm$ 0.71 \\
Ap15 & -               &    -             & 0.65 $\pm$ 0.12 & 2.60 $\pm$ 0.19 \\

\hline
\end{tabular}
\end{table}

The results point to extinction variations up to a factor of 4 within
the NLR. However, the E(B-V) derived from the hydrogen ratios suffer from
two major problems: the Pa$\beta$ line may be contaminated with the 
[Fe\,{\sc ii}]~1.278~$\micron$, leading to an overestimation 
of the reddening determined from Pa$\beta$/Pa$\gamma$. Also, the Pa$\gamma$ 
line is in the wing of the much stronger He\,{\sc i}~1.083~$\micron$ line. 
Therefore, problems with the deblending of these two lines could lead 
to larger uncertainties in this calculation. Note that
the uncertainties quoted are determined only from the S/N around the region
containing the lines of interest and they do not take into account 
potential deblending issues.

Differences between the reddening obtained from the [Fe\,{\sc ii}]  lines
and hydrogen lines are expected, since these lines are probably not
emitted from the same region. Keep in mind that the uncertainties in the
E(B-V) from the [Fe\,{\sc ii}] lines can also be large
since these lines are intrinsically weak. 

The analysis of Kraemer et al. (1998) indicates that there may be some 
dust mixed with gas in various amounts in this galaxy. Veilleux et al. (1997)
obtained a value of E(B-V) = 0.34 using the Balmer decrement. Our average
result for the NLR from both determinations is above this value. This might be
an indication that the NIR might come from dustier, deeper regions of the NLR.

The extinction obtained from the two line ratios for the
various apertures are shown in Figure 11.
The nuclear region seems to be
more dusty than the adjacent regions but secondary peaks of extinction
are also observed, with maximums around 150 - 200 pc from each side of the nucleus.
These positions are also close to the region where the second component of the 
emission lines has a maximum, and may be associated to the presence of dense
clouds mixed with dust in this region.

\begin{figure} 
\includegraphics [width=85mm]{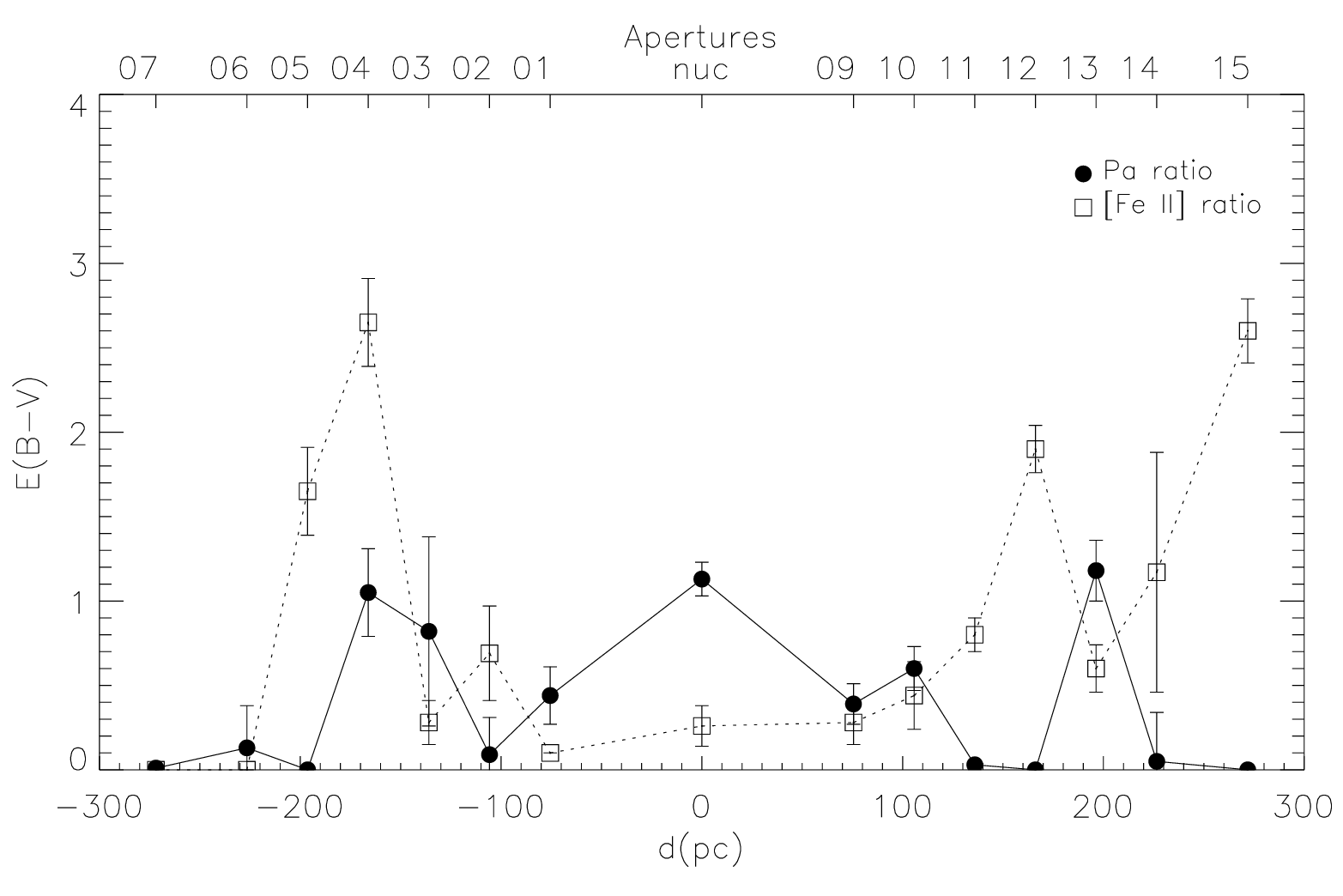}
\caption{E(B-V) derived from emission lines as a function of distance from the center.
Negative distances represent the north direction and positive distances the south.}
\end{figure}

\subsection{H$_2$ and [Fe II] lines}

Emission from H$_2$ molecules are observed in most apertures of NGC~1068,
and has been the subject of extensive analysis by different authors. 
Reunanen et al. (2003) suggest a star formation origin for these lines as a result of UV heating
from OB stars. Whether the heating mechanism is thermal or fluorescence
is not clear because neither the 2-1~S(1)~2.247$\mu$m nor the 1-0~S(0)~2.223$\mu$m 
emission were detected in the 
nucleus. They derived an upper limit of 5300~K for the vibrational temperature in the inner
few hundred of parsecs of this object.

We used their method to estimate the mass of the hot molecular gas present 
in each aperture. Assuming T = 2000K, a transition probability A$_S(1)$ = 3.47 x 10$^{-7}$ s$^{-1}$
(Turner et al. 1977) and the population fraction in the $\nu$ = 1, J=3 level
f$_{\nu = 1,J=3}$ = 0.0122 (Scoville et al. 1982),
the H$_2$ mass was obtained from

\begin{equation}
m_{H_2} \approx 5.0875 x 10^{13} D^2 I_{1-0S(1)} 10^{0.4277A_{22}}
\end{equation}

 where D is the 
distance in Mpc, I$_{1-0S(1)}$ is the observed flux of H$_2$ 
$\lambda$2.121 $\micron$ in erg cm$^{-2}$ s$^{-1}$ and A$_{22}$ is the 2.2-$\micron$
extinction calculated through the E(B-V) from Table 4.
The mass m$_{H_2}$ is given in M$_{\sun}$.

\begin{table}
\caption{Mass of the excited molecular hydrogen}
\begin{center}
\begin {tabular} {cccc}
\hline
 South &    m$_{H_2}$ [M$_{\sun}$] & North &  m$_{H_2}$ [M$_{\sun}$]   \\
\hline
Ap08 & 11.8 & Ap16 & -- \\
Ap07 & 13.5 & Ap15 & 49.8\\ 
Ap06 & 16.8 & Ap14 & 15.1\\
Ap05 & 30.9 & Ap13 & 189.5\\
Ap04 & 82.0 & Ap12 & 65.7\\
Ap03 & 31.4 & Ap11 & 52.0\\
Ap02 & 48.7 & Ap10 & 72.6\\
Ap01 & 78.1 & Ap09 & 98.2\\
\hline
Nuc  & 449.8& &\\

\hline

\end{tabular}
\end{center}
\end{table}

The results for each aperture are given in Table 5. We have to point here
that it is now well known that the distribution of H$_2$ (perpendicular to the NLR)
is so that we miss a large fraction of it in our slit (Sanchez et al. 2009, Galliano
et al. 2003).
These values are in general agreement with the values obtained by 
Reunanen et al. (2003) for Seyfert 1 and 2 galaxies, although the nuclear 
value we obtained is much smaller than what they obtained for NGC~1068.
Several reasons may be invoked to explain the difference.
The aperture they use to do the analysis
is much larger than ours, encompassing all our apertures. As can be seen in 
Table 4, the extinction varies and the higher values are not in the center.
Reunanen et al. probably got an average of these values. 
In addition, they obtained A$_{22}$ 
from the continuum
extinction, assuming that the difference for the line extinction would
not be large. This may be true for many Seyfert galaxies, but the continuum
of the nuclear region of NGC~1068 seems to be extremely dusty, as can be seen in Figure~7, by
the inclination of the spectrum.

Using the extinction we derived from the emission lines (section 4.2) we
obtained a much
smaller value for the hot molecular gas mass. 

\subsection{Excitation mechanisms of the H$_2$ and [Fe II] lines}

Galliano et al. (2003)  analyze the possibility of explaining
their observations of the MIR emission with diffuse dust
in the ionization cone. This model is discarded because the
diffuse dust would shield the central engine as well as the
BLR from any direction, requiring an ionizing mechanism
other than photoionization to explain the NLR emission.
On the other hand, they favor a model of thick dust clouds
in the cone, with a small covering factor, heated by
the central source radiation, without discarding additional
heating mechanisms by  shocks generated by the interaction of
radio jets and gas clouds.

In order to explain the observed molecular emission  perpendicular
to the cone axis, Galliano et al.(2003) built  a numerical model
combining a molecular disc and an axisymmetrical, no coplanar,
X-ray absorber (see their Figure 9), reached by X-rays from the
central source. This two component model reproduces quite well
the main features observed in the  H2 and CO emission maps.
 
More recently, the excitation of near infrared H$_2$ and Fe II lines have been 
discussed by Rodr\'{\i}guez-Ardila et al. (2004) based on observations covering 
the inner 300 pc of most Seyfert galaxies and minimizing contamination 
by the host galaxy. They showed that the diagnostic diagram H$_2$$\lambda$2.121/Br$\gamma$ 
versus [FeII]$\lambda$1.257 /Pa$\beta$ proposed by Larkin et al. (1998) is an efficient 
tool to separate objects by the level of nuclear activity, i.e., AGN, Starburst or Liner. 
By including more objects, Rodr\'{\i}guez-Ardila, Riffel \& Pastoriza (2005) further test the 
efficiency of the diagram. For NGC~1068, we obtained 
0.7 $<$ H$_2$$\lambda$2.121/Br$\gamma$ $<$ 1.3 and 0.6 $\leq$ 
[FeII]$\lambda$1.257 /Pa$\beta$ $\leq$ 2.8. Our results are shown 
in Figure 12 (filled circles) together with those observed by Rodr\'{\i}guez-Ardila 
2004 and 2005 (crosses); the dashed lines show the limits suggested by these authors. 
The observational data for NGC 1068 corresponding to the nucleus, as well as to different 
locations along the slit, are in the AGN region of the diagram. This reinforces the results 
found previously, i.e., that this diagram is a helpful tool for classifying objects according 
to their degree of excitation, and shows that line ratio intensities from slits at different 
distances from the center can be also used.

In addition,
the [FeII]$\lambda$1.257/[PII]$\lambda$1.188 line ratio can be used 
as a tracer of the ionization mechanism of the emitting gas and depends only on the 
Fe/P abundance. 
Following Oliva et al. (2001), in photoionized gas, like HII regions 
and planetary nebulae, this line ratio is less than 2, while in shock dominated gas its 
value is higher than 20. 
The high value would be due to the evaporation of the iron-based grains in the shock 
front, increasing the Fe abundance relative to P in the low-ionization region. 
Our data show 0.70 $\leq$ [Fe II]/[P II] $\leq$ 3.0. Such low values might 
indicate that, either like we mention in section 3.1, the high value of [PII] observed here
might be due to an overabundance of phosphorus, or
most of the Fe is locked in grains and the emitting gas is photoionized.

\begin{figure} 
\includegraphics [width=87mm]{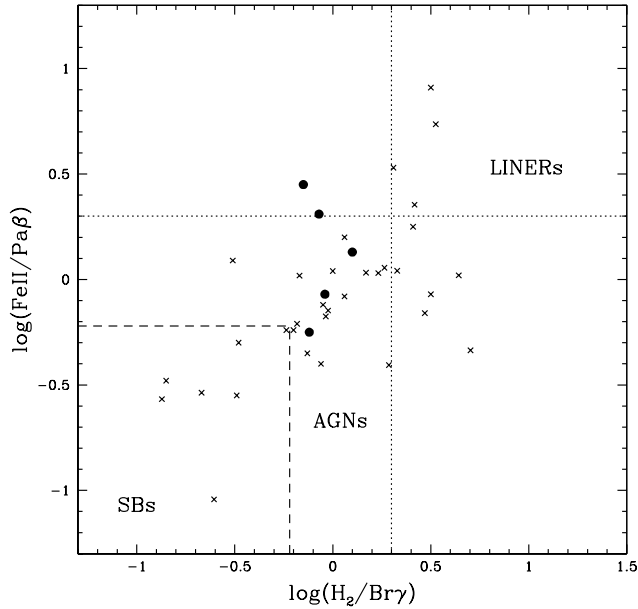}
\caption{Ratios H$_2$ 2.121/Br$\gamma$ versus [Fe II] 1.257 /Pa$\beta$. 
Filled circles represent our data for NGC 1068. Crosses to 
data obtained by Rodr\'{\i}guez-Ardila et al. (2004, 2005).}
\end{figure}

\section{Physical Conditions of the gas:  NIR coronal and low-ionization emission-lines}

In the last few years, photoionization models have been largely used to 
analyze the physical conditions of the NLR of NGC~1068. Some of these 
studies tried to reproduce the observed emission-line intensities in a
 wide wavelength range (Alexander et al. 2000, Kraemer \& Crenshaw 2000a,
 Kraemer \& Crenshaw 2000b, Spinoglio et al. 2005), while others  were
 based on selected lines (Groves et al. 2004, Groves et al. 2006, Martins, 
Viegas \& Gruenwald 2003, Rodr\'{\i}guez-Ardila et al. 2006). Recently, Osaki (2009) 
focus his analysis on the kinematic and excitation structure of the NLR of NGC~1068.

Although the basic photoionization codes used in the literature are very similar, 
the choice of the input parameters by different authors may change - in particular, 
the choice of the filling factor. Since one-dimensional photoionization codes are 
unable to properly deal with an inhomogeneous gas distribution, the optical depth 
at each cloud shell is multiplied by a chosen filling factor (see, for instance, 
Osterbrock 1989). The net effect is to increase the size of the low ionized zone 
as well as to decrease the average temperature of the cloud, which could mislead 
the conclusions. Following Martins et al. (2003), in the present paper we adopt a 
multicloud model which should provide a better simulation of the inhomogeneous NLR of NGC 1068.

Another important issue concerning photoionization models is the use of the 
so-called ionization parameter U to characterize the intensity of the ionizing 
radiation continuum. In fact, U is the ratio of the density of the ionizing 
photons reaching the cloud to the gas density. It is a convenient free parameter 
when using diagnostic diagrams to analyze emission-line ratios coming from different 
galaxies. However, in photoionization models for a given NLR where the distance D 
from the ionizing radiation source as well as the source luminosity are known, U is 
not a free parameter anymore but it is constrained by the source luminosity, the 
distance from the ionizing radiation source and the gas density (Ferguson et al. 1997). 

As shown in Tables 1, 2 and 3, the main emission line intensities were obtained as a 
function of the distance to the center. These measured line intensities result from 
the contribution of all the clouds in the line of sight covered by each aperture. 
It is not in the scope of this paper to provide a specific model for the observed 
data from each aperture. 
Our objective here is to give a scenario to try to explain the 
observations in general.
Thus, in the following, we search for photoionization models which can reproduce 
the observed range of intensities.

\subsection{Multi-cloud photoionization model}

First attempts to model the emission-line region with a homogeneous gas
photoionized by the nuclear radiation failed to reproduce the observed NIR
line intensities. 
As shown below, at least two density regions are necessary to explain 
the observed data. 

In the following, we present a two-cloud model for 
the NIR emission-lines using the photoionization code Aangaba 
(Gruenwald \& Viegas 1992). The model must reproduce the observed 
coronal lines [SIX]$\lambda$1.252, [SVIII]$\lambda$0.991, 
[SiX]$\lambda$1.430, and [S VI]$\lambda$1.962, as well as the 
[SIII]$\lambda\lambda$0.907, 0.953, and [SII]$\lambda\lambda$1.032 
emission lines (Tables 1, 2 and 3). 

The sulfur line intensity ratios are not directly dependent on the 
adopted chemical abundance. Thus, they are first used to limit the other 
input parameters, i. e., U (ionization parameter), n (density)
and D (distance from the ionizing source), given the ionizing radiation luminosity 
for NGC~1068, L$_{ion}$. The model results are then compared to other line 
ratios, including those for [FeXI]$\lambda$0.7889, [FeX]$\lambda$0.6374, 
[FeVII]$\lambda$0.6087 (Rodr\'{\i}guez-Ardila et al. 2006).  We assume a solar 
chemical abundance (Grevesse \& Anders 1989).

For NGC 1068, the ionizing radiation luminosity is about 10$^{44.5}$ erg/s 
assuming an UV power-law index equal to -1.7 and an X-ray index of -0.5 
(Pier et al. 1994, if the reflection factor  is 0.01).

In order to fit the high-ionization lines, a large value of U is usually 
required. However, due to the relationship between U, n, R and L$_{ion}$, 
a high U requires a low n, and a low distance from the ionizing source. In 
the models presented by the different authors referred above, the density 
value varies from 10$^2$ cm$^{-3}$ (Groves et al. 2006) to 10$^6$ cm$^{-3}$ 
(Ozaki 2009), although some of them use filling factors as low as 6.5$\times$10$^{-4}$.

From the observed coronal emission line intensities we  note that for gas 
clouds extending to 200 pc, values of U~$>$~0 are required to fit high-ionization 
lines, and could be reached only for densities lower than 10$^2$ cm$^{-3}$ 
(Rodr\'{\i}guez-Ardila et al. 2006, figure 12). After running a series of models 
for a homogeneous low density gas around the ionizing radiation source, we 
selected n = 1 cm$^{-3}$ and U = 10 to better fitting the coronal lines; 
such a model is called model A. The line intensities relative to H$\beta$ 
obtained for model A, for an optically thick cloud ($\tau^A_{Ly} \rightarrow\infty$), 
are given on the second column of Table 6. However, such a model also produces a 
large semi-ionized region, resulting in a [SIII]/[S II] line ratio too high. 
Thus the final model for the low density gas should be matter bound (see below).
 
While the coronal line ratios are more sensitive to the U parameter, 
the [SIII]/[SII] line intensity ratio is more dependent on the gas density. 
The observed values indicate that densities 10$^6$ cm$^{-3}$ or higher are 
required to reproduced the observed line ratio. Notice that high density clouds 
(10$^{5.3}$ cm$^{-3}$) are also favored by Ozaki (2009) when modeling the structure 
of the NLR of NGC 1068 using the optical emission-lines [FeVII], [OIII], and [OI], 
relative to H${\beta}$. Thus, we assume that high density clouds (10$^6$~cm$^{-3}$) 
are radiation bound, located farther from the central source and being
shielded from the UV, but not from the X-ray, by the low density gas producing the coronal 
lines. The results for a model using an ionizing spectrum with E $\geq$ 200~eV are 
shown in Table 6 (model B). In addition to [SIII] and [SII] lines, this high density 
gas produces also the [SiVI] emission line, but not the other coronal lines as expected. 

The high density clouds contribute mainly to the low ionization emission lines. 
In order to obtain the fraction f of the contribution of the high density clouds 
relative to the coronal gas, we use the [SVIII]/[SIII] observed line ratios 
(Table 1, 2 and 3). Since [SIII] comes both from models A and B, such line ratio 
depends on the optical depth at the Lyman limit of the low density region. As 
noticed above, the [SiVI] line could partially come from the high density clouds. 
However their contribution is much smaller than that from the low density region, 
so that the [SiX]/[SiIV] ratio is practically independent of f. As said above, 
the low density cloud must be matter bound for not emitting too much [SII] or [SIII]. 
The best results are obtained with $\tau^A_{Ly}$ for model A between 1 and 5.  
In Table 6, the line intensity ratios for $\tau^A_{Ly}$  = 5 and 1 are listed 
(third and fourth columns, respectively). For $\tau^A_{Ly}$ of model A equal 1 
and 5, the corresponding fractions f are equal to 0.078 and 0.065, respectively, 
leading to a luminosity ratio of H${\beta}$ between the two components equal to 
10.2 and 38.7. The range of the observed infrared line ratios is shown in Table 7, 
compared to the results of the two possible two-cloud models, i. e., the combination 
of model B with model A (for  $\tau^A_{Ly}$ =1 and 5). Notice that detailed models 
for each aperture may require a range of densities. 

\begin{table}
\caption{ Line intensities for components A and B, relative to H$\beta$}
\begin {tabular} {lcccc}
\hline

  & mod A & mod A & mod A & mod B \\
  & $\tau^A_{Ly} \rightarrow\infty$ & $\tau^A_{Ly}$=5 & $\tau^A_{Ly}$=1 & $\tau^B_{Ly} \rightarrow\infty$ \\ 
\hline 
$[$S II$]/H\beta $   & 3.5E-3   & 6.2E-5   & 9.2E-6 & 7.6E+0 \\ 
$[$S III$]/H\beta$   & 9.1E-1   & 4.2E-1   & 9.1E-2 & 7.1E+1 \\
$[$S VIII$]/H\beta$  & 3.2E-2   & 4.5E-2   & 1.4E-1 & 0.0E+0 \\
$[$S IX$]/H\beta$    & 7.9E-2   & 1.1E-1   & 3.5E-1 & 0.0E+0 \\
$[$Si VI$]/H\beta$   & 6.6E-2   & 9.1E-2   & 2.5E-1 & 4.2E-5 \\
$[$Si X$]/H\beta $   & 2.1E-2   & 2.9E-2   & 9.0E-2 & 0.0E+0 \\
$[$Fe VII$]/H\beta$  & 1.1E-1   & 1.5E-1   & 4.5E-1 & 0.0E+0 \\
$[$Fe X$]/H\beta   $ & 4.9E-2   & 6.8E-2   & 2.2E-1 & 0.0E+0 \\
$[$Fe XI$]/H\beta  $ & 2.2E-2   & 3.1E-2   & 9.7E-2 & 0.0E+0 \\

\hline
\end{tabular}
\end{table}

\begin{table}
\caption{ Results for the multi-cloud model (in logarithm)}
\begin {tabular} {lccc}
\hline
& observation & multi-cloud model & \\
& & $\tau^A_{Ly}$=1 & $\tau^A_{Ly}$=5 \\
\hline
$[$S III$]$/S$[$ II$]$   & +0.8 to +1.2 & +1.0 & +1.1 \\
$[$S VIII$]$/$[$S III$]$ & -2.0 to -1.5 & -1.7 & -1.7 \\ 
$[$S IX$]$/$[$S VIII$]$  & +0.0 to +0.7 & +0.4 & +0.4 \\ 
$[$Si X$]$/$[$Si VI$]$   & -0.4 to +0.0 & -0.4 & -0.5 \\
$[$Fe X$]$/$[$Fe VII$]$  & -0.5 to -0 4 & -0.3 & -0.3 \\ 
$[$Fe XI$]$/$[$Fe X$]$   & -0.9 to -0.2 & -0.4 & -0.4 \\
\hline
\end{tabular}
\end{table}

All line ratios are in the range of the observational data 
(Tables 1, 2 and 3), including the coronal iron lines from Rodr\'{\i}guez-Ardila et al. 
(2006). In that paper, only models with density $\geq$ 10$^3$ cm$^{-3}$  were
used, and pure photoionization models could not reproduce the line ratios.
Here we relax this lower limit, and most of the ratios could be reproduced.
This, however, does not discard the possibility of having other ionization mechanisms
playing a role. To put a stronger constrain on this issue it would be
necessary to fit a large amount of emission-lines in a broader 
range of wavelengths, which is out of the scope of this paper.

\section {The Central Mass}

We have used the stellar absorption features to estimate the 
mass of the central region inside the extraction apertures.
The velocity dispersion was estimated using the same velocity 
distribution and methodology proposed 
by Van der Marel \& Franx (1993). The $h3$ and $h4$ coefficients, 
which measure the distortions from a gaussian distribution, are very 
uncertain and therefore we restrict our discussion to the velocity 
dispersions. The results were obtained using as template a mixture of 
standard stellar spectra with the same resolution as our nuclear spectra 
of NGC1068 and high S/N. The spectral region was the 
same used by Tamura et al. (1991) for this same galaxy, around 2.26 $\micron$, 
where there are strong molecular absorptions due to CO. This region seems 
adequate since the interstellar absorption is small and there are no emission lines.
Errors in the velocity dispersions were obtained using the S/N of the spectra.
An example of the fit is shown in Figure 13.

\begin{figure} 
\includegraphics [width=80mm]{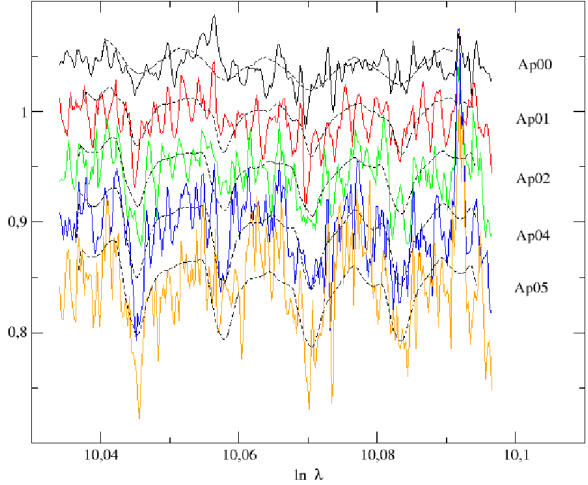}
\caption{Example of the fitting process. Color lines represent the 
observed spectra and dashed black lines are the ajusted model.}
\end{figure}

The results are shown in Figure 14 and we can observe that the velocity dispersion 
rises strongly in the central region, probably due to the AGN contribution. 
Moreover in this very central region the light contribution of the AGN and the 
hot dust contaminates 
the stellar continuum and makes it difficult to measure the velocity dispersion. Away 
from the center the nuclear contamination is much lower in our extracted spectra and 
we observe that the velocity dispersion remains quite large for a typical spiral galaxy. 
Nevertheless moving outwards to 100 pc from the nucleus the velocity dispersion is quite 
close to the values already reported in the literature. 

Using these determinations and the Virial theorem we give a very crude estimate showing 
how the central mass depends on the adopted nuclear distance. Our approach consists in using a 
simple model that can be resolved analytically and compared to the data. Basically 
the model consists of an stellar core with uniform density, $\rho_*$, extending to a 
maximum radial distance $R_*$. The total mass of this stellar core is $M_*$ and in 
the central region we add a punctual mass, M$_0$, that simulates the presence of the 
AGN and its accretion disk. The total mass profile can be readily estimated as

\begin{equation}
M(R)= M_* \left(\frac{R}{R_*}\right)^3+M_0
\end{equation}

\noindent where $R$ represents the radial distance. Assuming that the 
velocity distribution of the stellar component is random and isotropic we derive that 

\begin{equation}
 {\sigma^2}_V(R) \approx GM_* \left (\frac{1}{5} \frac{R^2}{R_*^3} + \frac{\gamma}{3} \frac{1}{R} \right)
\end{equation}

\noindent where $\sigma^2_V$(R) is the velocity dispersion and 
$\gamma$ = M$_0$/M$_*$ is a measure of the relative importance of the 
central mass compared with the core stellar distribution. However we only 
have access to the average kinetic energy of the stars projected in the line 
of sight of the observer. Weighting this component by its luminosity contribution, 
or by mass for a constant mass-luminosity ratio, we have

\begin{equation}
 {\sigma^2}_{v,obs}(r) = \frac {\int_0^L \sigma^2_v(R) \rho(R) dl}{\int_0^L \rho(R)dl}
\end{equation}

\noindent and assuming that the density of the stellar core is uniform, 
we finally obtain our estimation for the observed line of sight velocity 
dispersion as

\begin{eqnarray}
 \lefteqn{{\sigma^2}_{v,obs}(r) =  \frac{\sigma^2_{v,obs}(R_*)}{3+5\gamma}}\\ \nonumber 
& &  \left [ 1 + 2 \frac{r^2}{R_*^2}
+  \frac {5\gamma}{(1-r^2/R_*^2)^{1/2}} sinh^{-1}\left (\frac{R_*^2}{r^2} -1 \right)^{1/2} \right]
\end{eqnarray}

For large values of $\gamma$ the central mass dominates and we simply have a keplerian 
profile which is shown as the blue continuous curve in Figure 14. We can see that the 
observed profile is completely inconsistent using the usual BH mass estimation 
($\simeq 1 - 2 {\times} 10^{7} M_\odot$) found in the literature (Lodato \& Bertin 2002 ). To explain 
the high velocity dispersion of  255 km/s in the central region we need to include an 
stellar core of $1.5 {\times} 10^{10} M_\odot$  extending to approximately  200 pc 
from the nucleus shown as the dotted line. Therefore a value $\gamma= 6.6 {\times} 10^{-4}$ 
would be a more appropriated contribution of the central mass.

\begin{figure} 
\includegraphics [width=80mm]{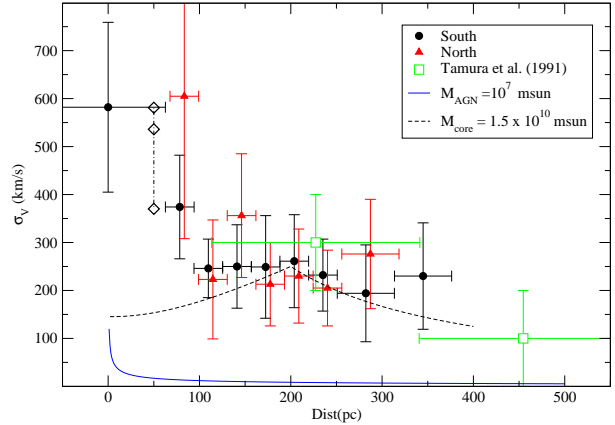}
\caption{Core model applied to NGC 1068}
\end{figure}

\section{Summary and conclusions}

We carried out long-slit spectroscopy in the Near-IR of NGC~1068
and presented here for the first time the
whole range from 0.9 to 2.4$\micron$, extending 15" in the nuclear 
region. The nuclear emission shows many important emission lines, from low ionization
to coronal lines, molecular lines and absorption features. 
We show that many emission lines are composed by two narrow 
components, one in the galaxy rest frame, and the other blueshifted in the
south and redshifted in the north apertures. The relative intensity of the 
peaks also varies and they seem to be correlated with the jet of the galaxy.
No broad component ($\sim$~2000) was identified in any line profile. 
Not all emission lines have double peaked profiles and some lines are present only
near the center, particularly the high ionization lines. This is a clear indication
that the emitting gas is more ionized in the nuclear region.

We estimated the extinction based on Paschen hydrogen and
[FeII] emission lines. E(B-V) is highest in the nucleus then in the
adjacent regions, but there are peaks around 150 - 200 pc from
each side of the nucleus. This is also close to the region where
the second components of the emission lines have a maximum, and 
may be associated with dense clouds around these regions.
H$_2$ emission is very extended, but it is
difficult to identify the ionization process because only the 1-0(S1) line could
be measured.

Simple photoionization models were used to try to explain the emission lines,
and a first approximation shows that the line ratios can be explained with
pure photoionization, as long as low density clouds are present.
A more detailed analysis however is necessary, to rule out other ionization
mechanisms. 

We also used the stellar signatures to determine the mass of the central
region, obtaining a stellar core of 1.5$\times$10$^{10}$M$_{\sun}$.

\section*{Acknowledgments}
This research has been partially supported by the Brazilian agency FAPESP (2007/04316-1).
ARA acknowledges partial support
of CNPq through grant 308877/2009-8.
The authors would like to thank Dr. Sueli Viegas for the helpful 
insights and comments on the paper. We also thank the anonymous referee for valuable comments.
The IRS was a collaborative venture between Cornell 
University and Ball Aerospace Corporation funded by 
NASA through the Jet Propulsion Laboratory and Ames Research Center.

\bsp

\label{lastpage}

\end{document}